\documentclass{emulateapj}
\newcommand{\xs}{XSPEC{}}

\newcommand{\isis}{ISIS{}}
\newcommand{\slirp}{SLIRP{}}
\newcommand{\slang}{S-Lang{}}
\newcommand{\tcl}{TCL{}}
\newcommand{\fort}{{Fortran}{}}
\newcommand{\initp}{{\tt initpackage}{}}
\newcommand{\bbp}{{Black Box Problem}{}}
\newcommand{\pgplot}{{\tt PGPLOT}{}}
\newcommand{\swig}{{\tt SWIG}{}}
\newcommand{\volview}{{\tt volview}{}}
\newcommand{\evtimg}{{\tt evt2img}{}}
\newcommand{\ftf}{{\em file $\rightarrow$ tool $\rightarrow$ file}{}}
\newcommand{\vwhere}{{\tt VWhere}{}}
\newcommand{\fcalc}{{\tt fcalc}{}}
\newcommand{\fv}{{\tt fv}{}}
\newcommand{\bfi}[1]{{#1}}

\shortauthors{Noble et al.}

\begin{document}


\title{Beyond \xs: Towards Highly Configurable Astrophysical Analysis}
\shorttitle{Beyond \xs: Towards Highly Configurable Astrophysical Analysis}


\author{M. S. Noble, M. A. Nowak}
\affil{Kavli Institute for Astrophysics and Space Research\\
	Massachusetts Institute of Technology\\
	Cambridge, MA 02139}
\email{mnoble@space.mit.edu}


\begin{abstract}

We present a quantitative comparison between software features of the
defacto standard X-ray spectral analysis tool, \xs\, and \isis\, the
Interactive Spectral Interpretation System. Our emphasis is on
customized analysis, with \isis\ offered as a strong example of
configurable software.  While noting that
\xs\ has been of immense value to astronomers, and that its
scientific core is moderately extensible---most commonly via the
inclusion of user contributed ``local models"---we identify a series of
limitations with its use beyond conventional spectral modeling.  We
argue that from the viewpoint of the astronomical user, the \xs\
internal structure presents a \bbp, with many of its important
features \bfi{hidden} from the top-level interface, thus \bfi{discouraging}
user customization. Drawing from examples in custom modeling, numerical
analysis, parallel computation, visualization, data management, and
automated code generation, we show how a \bfi{numerically} scriptable,
modular, and extensible analysis platform such as \isis\ facilitates
many forms of advanced astrophysical inquiry.

\end{abstract}


%\keywords{methods: data analysis --- methods: statistical --- X-rays: general}



\section{Introduction}
\label{intro}
The pursuit of science pushes not only the boundaries of knowledge, but
also the limits of technology used to acquire and analyze data from which
new knowledge can be distilled.  In this sense software systems for
scientific analysis are never truly complete, so it is crucial that they
be extensible.  This enables scientists to incorporate custom, experimental
codes into
the system and drive it in directions not yet supported, or perhaps even
envisioned, by its creators.  A corollary is that the mechanisms for
and results of the customization should be made as simple, powerful,
and general as possible, freeing the practitioner to concentrate more
on algorithmic or scientific concerns and less on the plumbing details
of a given computational platform.

This paper compares the extent to which two open-source spectral analysis
applications actively used in astronomy,  \xs\ \citep{1996ASPC..101...17A}
and \isis\ \citep{2002hrxs.confE..17H}, meet these general criteria.
The features of both systems are broadly surveyed, with contrasts
drawn from a series of examples representative of \bfi{astronomy in
practice}.  A heavy emphasis is placed upon the roles of interpreted arrays,
and the capability to quickly generate and numerically manipulate them
as powerful enablers of analysis.
The emergent theme is that numerical and architectural discontinuities
between the internals of \xs\ and its controlling \tcl\ interpreter,
referred to in the large as the \bbp, limit its full exploitation for
scientific tasks that are \bfi{within the reach of more configurable}
interactive analysis systems.  We demonstrate the strong modularity
and extensible scriptability of \isis\, and suggest that such open and
highly configurable systems offer deeper and wider promise for meeting
the needs of exploratory scientific analysis.

The latest major versions of \isis\ and \xs, {\tt 1.4.x} and {\tt 12.x}
respectively, were used in our study.  While \xs\ 11 remains in wide
use (the first public, non-beta release of \xs\ 12 was
circa 2005), it is more difficult to extend with a variety of custom
models\footnote{For example, \xs\ 11 supports only local models coded
in single precision \fort 77.}
and so in terms of extensibility would compare less favorably than does
\xs\ 12.

Guiding our endeavor will be the sense of what a typically motivated
scientist can achieve in each system with reasonable effort.  
On the premise that a strong programmer can eventually replicate
virtually any result using nearly any combination of languages or tools,
we are not concerned with what might in principle be achieved {\em from}
either \xs\ or \isis\, but rather what can be demonstrated {\em within}
each.  By avoiding hypotheticals we aim to transform a potentially
open-ended debate into a tractable, concrete discussion.
Towards this end we define {\em working within an application} to mean
the execution of commands, or portions thereof, which only access
or modify the code or data within the applications addressable memory.
In this scheme, for example, issuing an operating system call from \isis\ to
spawn an external visualization tool would not be ``working in \isis,"
however enlarging its address space by importing a shared object module,
say to smooth an image in memory, would be.

We recognize the utility of integrating complementary but distinct
programs into a single logical entity.  Building on trust
in known tools, it is a sensible scheme for managing complexity, fostering
reuse, and balancing deadlines with goals.  Cooperating applications can,
with sufficient effort, be made to appear as seamless as a single
program.  \bfi{One} should not, however, equate the act of
\begin{itemize}
\item dumping internal state to disk files, spawning a series
of tools to read this state and create temporary data products
along the way, then loading these new data back into the parent
application
\end{itemize}
with the ability to
\begin{itemize}
\item invoke a function to operate upon a data structure, with both
resident in memory.
\end{itemize}
\bfi{In this paper we will argue that the latter offers greater immediacy
of use, as well as significantly more algorithmic, data, and visualization
flexibility.  It also addresses
real scalability concerns \label{scalability}} as datasets increase in
size or tasks increase in time complexity.  A stark example of this
is given in \cite{2005ASPC..347..444D}, which discusses the methods
employed to analyze the megasecond observation of Cassiopeia A
\citep{2004ApJ...615L.117H} by the Chandra X-Ray Observatory in 2004,
which yielded the first ever map of electron acceleration in a supernova
remnant \citep{2006NatPh...2..614S}.  Over 9 million PHA spectra
were extracted from over 300 million events, with hundreds of thousands
of spectral fits performed in \isis\, to yield arc-second resolution spatial
maps of plasma temperatures, line emission, and Doppler velocities.  Custom
modules were developed to manage the multi-gigabyte files involved and
distribute the spectral extractions
and fits across a network of workstations using the Parallel Virtual
Machine \citep{Geist:1994:PPV}.  Based upon the time
required to extract a single spectrum from these enormous datasets, with
the standard Chandra {\tt dmextract} tool in CIAO (ca. version 3.2), the
authors extrapolated that the spectral extractions alone would have
required nearly 10 years had a tool-based approach been taken.  In
contrast, each run through the entire PVM-based analysis sequence
required less than 12 days to complete.

\section{Feature Survey}
\label{feature_survey}
\xs\ and \isis\ have been primarily concerned with fitting models to 1D
spectra:
a theoretical model of photon counts is calculated, convolved with an
instrument response, and compared to the actual photon counts observed by
the detector, using a given fit statistic (typically $\chi^2$).  Each
provides mechanisms for indexed loading and management of data
(observed counts, instrument responses, et cetera), as well as the
capacity to visually inspect, using \pgplot\ to construct a variety
of 2D plots, the data, models, residuals, fit statistics, and so
forth.  While initially targeted for X-ray analysis, both \xs\ and \isis\
have been utilized in multiple wavebands.
The primary authors of \xs\ and \isis\ are practicing astronomers with
active publication records in the domains served by their applications.
Both applications are thus subject to continuous, rapid, and
rigorous scientific vetting. \isis\ and \xs\ update cycles are measured
in terms of days and weeks, which can be beneficial for users under time
constraints before their proprietary data go public.

\xs\ is the tool most widely used in X-ray astronomy for spectral fitting;
with a legacy spanning more than two decades and hundreds of citations,
its value to the scientific community is beyond question.  \xs\ bundles
more than 50 theoretical models, each of which
may also be used by external programs simply by linking to
its libraries.  The core set of models may be extended with either
precomputed file-based table models or compiled codes.  The bulk
of compiled \xs\ models are coded in \fort, however as part of a significant
redesign effort support for custom C and C++ models was recently added
(ca. 2005).
\xs\ was also one of the earliest astronomical applications to adopt a
widely-used, general-purpose scripting language for interactive use
and batch control; the incorporation of \tcl, introduced in version 10
and stabilized in version 11, provided programmability and a clear path
to graphical interfaces with the Tk widget set.

\isis\ was conceived to support analysis of high-resolution Chandra X-Ray
gratings spectra, then quickly grew into a general-purpose analysis system. 
\isis\ is essentially a superset of \xs, combining all of its models (though
they are not required for \isis\ operation) and more with the \slang\
scripting language \bfi{\citep{SLangRef}, whose mathematical performance rivals
both commercial packages such as MatLab \citep{Gilat.2004.MATLAB} or
IDL \citep{Kling.1999.IDL} as well as the numerical extensions for
popular open source languages \citep[\S \ref{numerics}]{Noble.2007.OpenMP}.}

As with \xs, users may define custom models in compiled code, external
tables, or as a string specifying simple arithmetic combinations of
existing models, but \isis\ takes
it further by also allowing models to be interpreted \slang\ functions;
this supports rapid prototyping and, because of the high-performance of
\slang\ numerics (\S \ref{numerics}), need not sacrifice speed for the
convenience of using an interpreter.  \xs\ does not support the use of
\tcl\ procedures as models.
% NB: mdefine command provides a crude approximation, but is not in XS 12
%and given that \tcl\ is poorly suited for numerical work (\S \ref{numerics})
%such may not even be advisable.

The chief means of controlling \xs\ and \isis\ interactively is by
textual prompt, with both providing VI and EMACS key bindings through
GNU readline, as well as filename completion and persistent command
history.  \xs\ accepts commands entered in abbreviated form, such as
{\tt mo} for {\tt model}, but does not auto-complete \tcl\ command or
variable names from partial input.  \isis\ provides strong auto-completion
facilities, allowing any \slang\ function or variable name to be
% NB: submit an ISIS patch to auto-complete . commands, too!
identified from minimal partial input.
In addition to the command prompt, optional packages providing
graphical interfaces, such as the
Gtk-based \vwhere\ (\S \ref{vwhere}) or the SAOds9 \citep{2003ASPC..295..489J}
image display module (\S \ref{ds9}), have been publicly available and
utilized within \isis\ for several years.
Graphical extensions for \xs\ have been discussed in the literature
\citep{1994ASPC...61...71J} and online, but, other than wholly external
tools like \fv\ \citep{1997ASPC..125..261P}, none seem publicly
available for use within \xs\ proper.

The \xs\ prompt presents an appealingly simple,
{\em everything-is-a-command} interface, and fosters ease of use by
offering brief, germane command names, assuming sensible defaults, and
in some cases transparently combining multiple operations into one: e.g.
loading both data and background files with a single command.  A typical
command is shorter and requires less punctuation in \xs, although the
auto-completion capability of \isis\ can compensate for its relative
verbosity.  In short, \xs\ makes it easy to perform many common spectral
analysis tasks.

\begin{figure}
  \begin{centering}
  \includegraphics[angle=-90,scale=0.3]{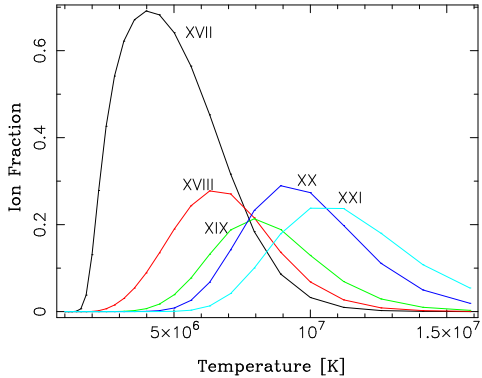}
  \caption{Ionization fractions for Fe XVII through XXI.}
  \label{ionfrac}
  \end{centering}
\end{figure}

The \isis\ prompt was designed to be a proxy \slang\ interpreter
(Fig. \ref{arch}), making \isis\ programmable and extensible since
inception.
Its {\em everything-is-a-function} philosophy permeates the implementation
and user experience, providing, at the cost of a steeper initial learning
curve, considerable flexibility and power.
Consider for illustration the Astrophysical Plasma Emission Database
(APED) of \cite{2001ApJ...556L..91S}.   Because \isis\ provides
programmatic access to virtually all of APED, curves like those of
Fig. \ref{ionfrac}, showing
the ionization fractions at 25 temperatures for Fe XVII through XXI,
can be created with just a few statements
{ \small
  \begin{verbatim}
     temperature = 10.0^[6.0:7.2:0.05];
     foreach ion ( [17, 18, 19, 20, 21] ) {
        fraction = ion_frac(Fe, ion, temperature);
        oplot (temperature, fraction);
     }
\end{verbatim}
}
\noindent
either directly from the \isis\ prompt or within analysis scripts.
The importance to X-ray analysis of this sort of facile manipulation of
atomic data is heavily underscored in
\cite{kallman-2007-79}.

\subsection{The Black Box Problem}
\bfi{
It is not clear how a similar ad-hoc query of APED could be formulated
in \xs, nor how the results of such might be used within it for further
analysis.  One problem is that the use of APED in \xs\ is highly constrained,
due to it being hidden, e.g., within the internals of the {\tt apec} model
and {\tt identify} commands.
Another obstacle is that high-level plotting in \xs\ is not generic: the
{\tt plot} and {\tt iplot} commands are also opaque, in that they take}
their vectors from internal state representing the current data or
model under inspection.  Intimately
tied to the semantics of 1D spectral analysis, these commands may not
be used until after a dataset has been loaded.

While \isis\ contains
similar visualization commands specifically tailored to spectral analysis,
the {\tt oplot} function shown above is part of a family of routines which
\bfi{allows one} to create arbitrary image, scatter, histogram,
and contour [over]plots from arbitrary data, independently of the current
spectra being analyzed.  This makes exploratory coding and visualization like
{\small
\begin{verbatim}
          isis>  x = [PI/2: 2*PI: PI/100]
          isis>  plot(x, cos(x))
          isis>  oplot(x, sin(x))
\end{verbatim}
}
\noindent
\bfi{as natural in \isis} as it is in other interactive analysis
systems such as IDL or PyRAF \citep{2001ASPC..238...59D}.  In \xs\ such
dynamic creation of interpreted arrays and direct mathematical manipulation
and visualization of them is more difficult, involving a melange of
\tcl\ code, \xs\ commands, intermediate files, and low-level QDP/PLT
directives.\footnote{Or going outside of \xs\ to use tools like \fv\
or XIMAGE.}
\begin{figure}[t]
  \begin{centering}
  \includegraphics[angle=-90,scale=0.30]{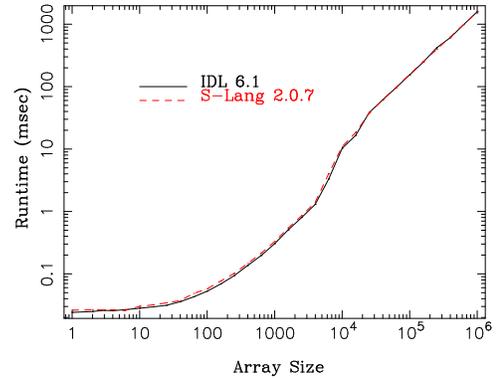}
  \caption{Performance of IDL 6.1 (binary distribution) and
        \slang\ (statically linked), on $\sqrt{b^2 - 4ac}$
        for arrays of various sizes.}
  \label{idl-plot}
  \end{centering}
\end{figure}
The issue is not that interpreted numeric arrays cannot be utilized in
\xs, because the \tcl\ extensions BLT and NAP (see next section) make
this possible; nor does the problem lie with \pgplot, since it provides
more capabilities than \xs\ currently exploits.  Rather, it is the \xs\
architecture which does not connect the two in a \bfi{straightforward manner,
leaving the user little choice but to look elsewhere for open-ended
numerics and visualization.  This opacity and disconnectedness is evident
in other functional areas of \xs, notably the manner in which data are
input \& managed (\S \ref{input}) and how custom models are specified
(\S \ref{custom-models}).
\begin{figure*}[t]
  \begin{centering}
    \includegraphics[scale=0.50]{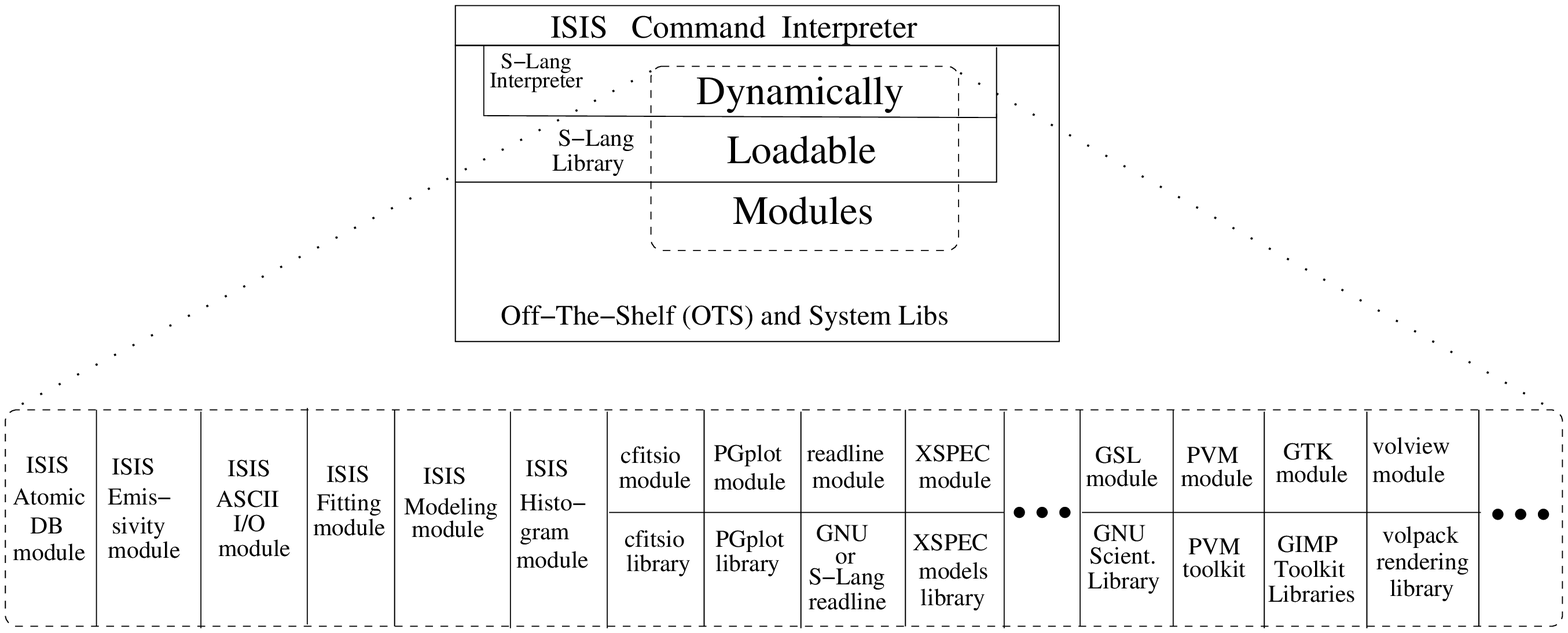}
    \caption{ The \isis\ { \tt main()} program is a thin ANSI C layer above
	the \slang\ interpreter, used primarily to establish hooks to
	modules and gather user input.  Taking modularity
	to this extreme, where carefully orthogonalized components provide
	the bulk of capability, provides a clean and flexible implementation.
	Most modules may also be
	used outside of \isis.}
  \label{arch}
  \end{centering}
\end{figure*}

Echoing similar criticisms levied upon graphical interfaces when compared
to command lines \citep{NextUIBrkThru,OscarCLI}, we collectively refer
to these issues as the \bbp: by hiding complexity to make common tasks
easy, uncommon or novel tasks can be made difficult or impossible.}

The ease of use that has been a hallmark of \xs\---written by astronomers
to do things astronomers need, in a way natural to them---has served the
community well.  A challenge, though, is that as instruments and the
data collected from them
grow in size and complexity, causing us to consider new questions and
possibly develop new techniques to answer them, the \bbp\ can engender
{\em You May Only!} patterns of analysis,
rather than foster rapid, ad hoc explorations of {\em What If?} scenarios.
More modular, configurable, and open implementations \citep{BeyondBlackBox}
can help resolve this tension, allowing applications to rapidly evolve to
suit specific user needs while freeing their primary authors of the
burden of coding and maintaining such enhancements themselves.

\section{High Performance Numerics}
\label{numerics}
Compact multidimensional numerics are a \bfi{major basis
for the popularity} of commercial toolsets such as IDL or MatLab.  The
innate capability of \slang\ in this regard was a primary motivator for its
adoption in \isis,\footnote{Another motivator was wide availability:
\slang\ is utilized in numerous open source projects and, in part by
virtue of being bundled as a core component of every major Linux 
distribution, is available on millions of machines worldwide.}
\bfi{and endows it with analytic expressiveness and
scripting performance not equaled in \xs}.

\begin{figure}[h]
  \begin{centering}
  \includegraphics[scale=0.80]{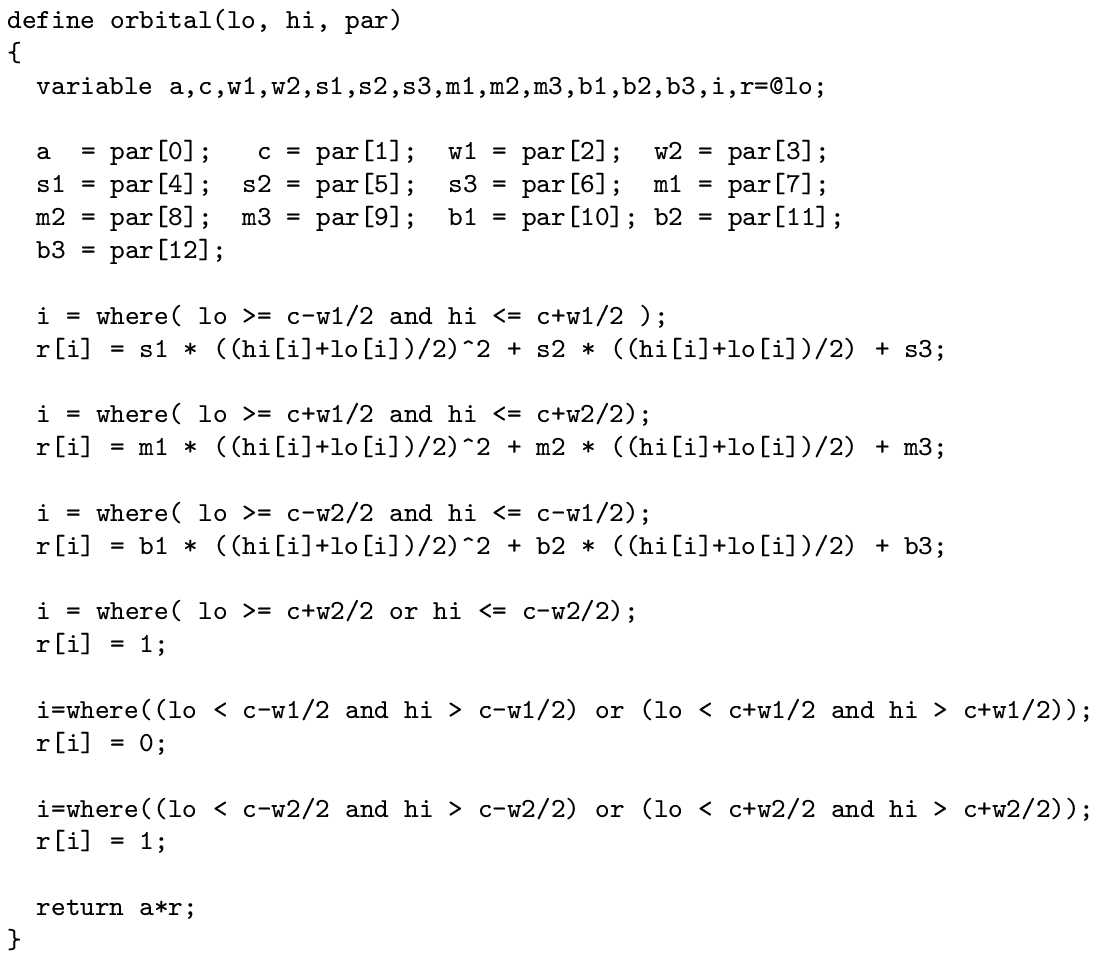}
  \caption{Pure \slang\ orbital model, fit
  by \isis\ directly to in-memory data arrays.
As described in \cite{TrowbridgeNowakWilms2007}, the model
represents a lightcurve of constant amplitude {\tt a}, with an eclipse
centered at {\tt c}, with a width {\tt w2} from the start of ingress
to the end of egress.  The ingress and egress have a quadratic form
({\tt m} and {\tt b} parameters), and there is a quadratic
``bounceback" ({\tt s} parameter) with width {\tt w1}.}
  \label{orbmodel}
  \end{centering}
\end{figure}

As indicated in the earlier code fragments, 
complex mathematical abstractions may be stated concisely in \slang,
without regard to whether they will operate upon scalars, vectors,
multidimensional arrays, or some combination of each, and are computed with
performance on par with commercial software (e.g., as in Fig. \ref{idl-plot}).
\bfi{In this section we discuss the more involved example of Fig.
\ref{orbmodel}, consisting of an orbital model implemented in pure \slang\
and taken ``as is" from an MIT research effort
The model was fit to a coarsely binned lightcurve (providing an example of
how \isis\ can model non-spectroscopic data directly from in-memory \slang\
arrays, as discussed in \S \ref{input}), and represents a steady amplitude
curve interrupted by an eclipse with a quadratic ingress and egress, with
low-level emission during the eclipse also modeled by a quadratic function.}
\citep{TrowbridgeNowakWilms2007}.

\begin{figure}[t]
  \begin{centering}
  \includegraphics[angle=-90,scale=0.3]{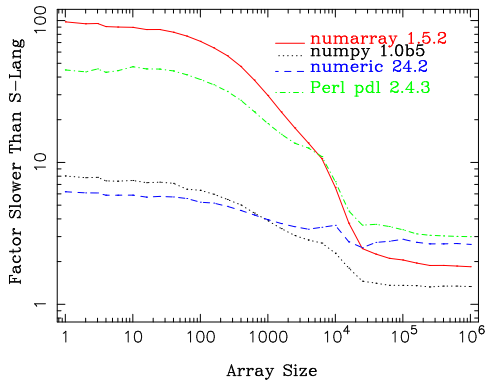}
  \caption{Performance of orbital models implemented in Perl 5.8.4
   and Python 2.4, relative to \slang\ 2.0.7.
   \tcl\ is not represented here because neither of its numerical 
   extensions, BLT and NAP, offered a clear equivalent to
   the {\tt where()} function.}
  \label{orbmodel-plot}
  \end{centering}
\end{figure}

As a means of assessing
the numerics and extensibility of \xs\ we attempted to translate this
model into \tcl; for completeness we also made Perl and Python translations,
since these languages are actively employed by astronomers in systems like
PDL and PyRAF and so are useful contrasts for gauging the numerical
capabilities of \slang.  Because neither \tcl\ , Perl, nor Python are
intrinsically vectorized we used numerical extension modules for each
conversion: BLT \& NAP for \tcl, PDL for Perl, and Numeric, NumArray,
\& NumPy for Python.
\noindent The Python and Perl conversions were straightforward, with
Python being the somewhat easier of the two; we did not complete the
conversion to \tcl, however, largely because neither BLT nor NAP provided
a clear equivalent to the {\tt where} command or a means of using the
results of such for array slicing.  This model would therefore need to be
recoded in a compiled language before it could be used in \xs.
A plot of the performance of Perl 5.8.4 and Python 2.4 implementations
of this model, divided by the corresponding \slang\ 2.0.7 performance,
is given in Fig. \ref{orbmodel-plot}.

The testing 
methodology behind Figs. \ref{idl-plot} and \ref{orbmodel-plot} is
described in \cite{Noble.2007.OpenMP}, along with memory statistics
and additional tests showing similar performance trends.
Briefly, the datapoints in each curve are the mean times of 1000
invocations of the model on a given grid size (31 in all, from 1 to 1e6
bins); all codes were compiled using GCC 3.3 with -O3 optimization, and
executed on a dual Athlon (1.8Ghz) machine running Debian 3.1.
Although not a comprehensive series of benchmarks, these results
\bfi{hint that the numerical engine of \isis\ is among the strongest
available.  High
performance scripting means that rapid development techniques---irrespective
of language---can be applied to a broad} scope of analysis problems,
\bfi{allowing the writing of compiled code to gradually become a last
resort instead of the primary avenue of attack}.
\slang\ bears a strong resemblance to C and IDL, arguably the most popular
scripting language used by astronomers, and in fact we and colleagues
have converted numerous IDL and MatLab scripts to \slang\ for use in 
\isis\, with \bfi{relative} ease.

\section{Array-Based Input}
\label{input}
As we found with the orbital model, converting such scripts to \tcl\ for
use in \xs\ would pose a \bfi{considerable challenge}: although
the primary interface of
\xs\ is a scriptable interpreter in which multidimensional arrays may
be created and mathematically manipulated---with the aid of the BLT or
NAP---the results of such cannot be straightforwardly utilized for
spectroscopy.  The fact that the internal data tables of \xs\ are not exposed
for direct population from the \tcl\ interpreter is another instance of
the \bbp; data may only be read from FITS files, and only via
the {\tt data} command or its INTEGRAL mission-specific variant
{\tt SPIdata}.  There is no documented provision by which interpreted
arrays may be used, for instance, to specify observed counts or
responses.

Reflecting a fundamental difference in \bfi{approach}, \xs\ is
more static than \isis, taking the position that extensive data
preparation happens outside the application.  An advantage of this
approach is self-documentation: well-written tools record the
path along which a FITS file travels as history keywords in its header.
A drawback is the underlying implication that input data need
little massaging during interactive analysis; when the need for such
data manipulation arises, the solution typically involves dumping
\xs\ state to disk files, and/or running FTOOLS or other programs
to generate new file products, then reloading these data back into \xs.

\bfi{This {\em files-only} data management paradigm can lead to slower
performance and more cumbersome analytics (\S \ref{scalability},
\S \ref{vwhere}).
It can also be an evolutionary disadvantage, i.e. by potentially limiting
the pool of individuals able or willing to contribute new I/O codes to \xs.}
As exemplified by the implementation of the {\tt SPIdata} command,
endowing \xs\ with the ability to directly read additional file or
mission formats requires detailed mastery of its internals.
This means general-purpose code generators like \swig\ \citep{1996SWIG}
or \slirp\ \citep{Noble.2007.OpenMP} cannot be leveraged to automate the
wrapping of I/O libraries and simplify the use of new formats in \xs. To
foster the widest possible use, by design such wrapper generators know
nothing about the applications in which the bindings they create will
be used.  They emit code targeted to the scope of a scripting language,
rather than application-specific C++ methods like
{\tt SPI\_Data::read()} which is suitable only for internal \xs\ use.

\subsection{Towards Dynamic Data Management}

\bfi{
\isis\ aims for a more dynamic and configurable approach to data management.
In addition to supporting file-based input in the manner of \xs, but with
ASCII in addition to FITS, \isis\ also permits most facets of the modeling
process to be specified directly from interpreted \slang\ arrays in memory,
including the theoretical and observed counts, ARF, RMF, and background.

The problem of augmenting spectroscopy with data in foreign formats
is thus reduced from having to master \xs\ internals to merely being
able to create \slang\ arrays from such files.  This is within the reach
of end users, not just application authors, particularly because automatic
code generators can then be employed to simplify the creation of scriptable
wrappers for the relevant I/O libraries.}

\bfi{As an example of how these considerations can matter in practical use,}
consider the HDF5 file format and I/O library \citep{1999IEEE...99...273F}.
While FITS is the standard format for archiving and distributing
astrophysical observations, HDF5 has become the defacto standard 
for storing astrophysical simulations such as those generated by
FLASH \citep{2001NuPhA.688..172F}.  Having the ability to easily
compare and contrast \bfi{observations with simulations, e.g., using
simulation output as source model input, could foster more sophisticated
analyses.  We have explored such questions in ISIS as part of the
HYDRA\footnote{http://space.mit.edu/hydra} project, using \slirp\ 
to generate the SLh5
module\footnote{http://space.mit.edu/cxc/software/slang/modules/slh5}
for reading and writing HDF5 data directly to and from in-memory \slang\
arrays.  The resulting objects may be sliced with \slang\ array manipulators
or mathematically transformed in arbitrary ways (\S \ref{numerics}) before
further analysis, such as being treated as fit data or model data, or
passed to 2D plotting or 3D volumetric routines for visualization.
}
This module also obviates
the need for a conversion tool to migrate data from HDF5 to FITS for
spectroscopy\footnote{http://www.hdfgroup.org/RFC/fits2h5/fits2h5.htm
describes preliminary efforts to go in the other direction.}.

This spirit of imposing fewer constraints on how its internal tables may
be populated or manipulated \bfi{leads to more flexible and mathematically
diverse analysis.  For example,}
consider the problem of grouping data to ensure an
adequate signal to noise ratio is obtained for each detector channel.
With \xs\ data are grouped prior to input, usually with the {\tt grppha}
tool. The assigned grouping persists for the life of the loaded dataset,
and may only be changed by re-running {\tt grppha} outside of \xs,
deleting the original dataset in \xs, and reloading it with the new
grouping.  With \isis\ spectra may be dynamically rebinned in memory---
using the functions {\tt rebin\_data()}, {\tt group\_data()}, or
variants---with the associated dataset remaining intact.  \isis\ also
provides options not available in the \xs\ {\tt grppha} paradigm, such
as grouping by signal to
noise ratio, or allowing datasets to be combined directly in memory,
bringing to bear the full numerical power discussed in \S \ref{numerics}
and \S \ref{beyond1D}, instead of shelling out to run tools such as
{\tt mathpha}, {\tt marfrmf}, or {\tt addrmf}---an approach of
more limited mathematical generality.
\isis\ also supports combining non-linear (e.g. piled-up) data, which
in general cannot be done with tools because of the model evaluation
involved.

\section{Custom Models}

\label{custom-models}
In \S \ref{numerics} we detailed how the performance of \slang\ couples
with the ability of \isis\ to use purely interpreted models to provide
an alternative to the traditional method of using compiled
codes as custom spectral models.  This significant feature, not
available to \xs\ users since it does not permit \tcl\ procedures to be
used as models, \bfi{can encourage rapid experimentation with short,
highly analytic codes.  These brief, self-contained scripts are
easily exchanged among users, as they may be loaded immediately into
\isis\, from any directory.   This is in contrast to code that
must be compiled into a shareable object then accessed from fixed
locations and in conjunction with external description files, such
as \$LOCAL\_MODEL\_DIRECTORY and lmodel.dat are respectively used in \xs.}

Moving to the realm of compiled models, while version 12 of \xs\ provides
many improvements over version 11---notably support for C and C++,
double precision numerics, and the \initp\ automated model wrapping
command---a number of limitations persist.
To begin, using local models at all means one must first build \xs\ from
source.  Although HEASOFT generally builds well, this can still be
daunting, and is unnecessary with \isis\ as both interpreted and compiled
user models may be used from a binary installation.
Next, model components in \xs\ must be designated as having a specific
form, e.g. additive, multiplicative, convolution, etc.  This too is 
unnecessary in \isis, as it places no restriction upon the computational
form of a model---any conceivable mathematical function may be used.

As \xs\ has been the defacto standard for so long, it is also easy
for spectral model developers to code expressly for \xs, by relying too
heavily upon its internal functions (such as {\tt udmget} for memory
allocation or {\tt xwrite} for message output) when crafting models.
When \xs\ was the only X-ray spectral modeling tool available this was
not a concern, but today it makes reusing models in other environments
more difficult.  For example, pulling the {\tt pexriv} model out of
\xs\ 11, which remains widely used, required
{\small
\begin{verbatim}
          -lxspec -lxspec_lfn -lxspec -lwcs
          -lxanlib -lreadline -lcfitsio -lpgplot
          -lcurses -ltermcap
\end{verbatim}
}
\noindent on the link line, plus the introduction of a ``dummy" function
to resolve an unused symbol name \citep{2004ApJ...609..972M}.
Standing out here is the recursive use of {\tt-lxspec},
and the need for auxiliary terminal management and plotting libraries
that provide no numerical or physics contribution to the model.
Although this situation is improved in \xs\ 12, these show that it is
not enough to simply make source code available
to the public: {\em orthogonality} is an important aspect of open and
flexible software.  To promote external reusability, code must also be
{\em separable} into logically distinct units free of unwanted
dependencies.  These are core implementation principles of \isis\
(Fig. \ref{arch}).

\subsection{Automated Model Wrapping}
While \initp\ does help automate the use of custom models in
\xs\ 12, it is not a generic, application-neutral wrapper generator in
the style of \slirp\ or \swig.  The result is that custom modeling cannot
be scripted in \xs\ with the same level of flexibility as in \isis.  For
example, \initp\ does not expose the routines it wraps to the top-level
\tcl\ interpreter, but rather makes them accessible only through the
the \xs\\ {\tt model} command.  Not being able to call
models directly from \tcl\ means the convenience of an interpreter
cannot be leveraged when testing and using the models.

This can be easy to overlook, because many model writers 
devise their own nominal tests in whatever compiled language they
used to write the model.  However, it would be valuable for both
writers and users of models if they could craft \tcl\ commands or
scripts to exercise them (in a more powerful and streamlined manner
than the multiplexing {\tt tclout} command, or faking a dataset) or
pass their output to other features in the application for further
numerical processing.  Consider Fortran common blocks, for instance,
and the \xs\ {\tt warmabs} model in particular:  after evaluation
the {\tt ewout} block in this model contains a list of the strongest
lines, sorted by element and ion.  Wrapping {\tt warmabs} with \slirp\
makes the content of this and every other common block in the model
accessible to \isis\ as \slang\ variables.\footnote{Common block values
may also be modified using regular \slang\ assignment
expressions.}  This provides scientists another means of accessing
the internals of a code, without needing to master the internals
of the application for which that code was initially written.

In addition, \initp\ wraps only the outermost interface routine of a
model: if the spectrum returned by a model is actually computed in
several steps by calling additional routines, those lower level routines
are not exposed by \initp\ for use in \xs\ outside the model evaluation.
As an example of how exposing the low level routines could be useful,
we consider
the {\tt pshock} and {\tt vpshock} models.  Internally, both rely upon
the lower-level {\tt ionseqs()} routine to calculate nonequilibrium
ionization ionic concentrations.  In response to a collaborator
(Ji 2007, priv. comm.) who wished to obtain these ionic
concentrations independently, we used \slirp\ to wrap {\tt ionseq.f}
in a matter of only minutes, which allowed {\tt ionseqs()} to be called
directly from a \slang\ script in \isis.

The jet model of \cite{2005ApJ...635.1203M} provides another good
example.  The output spectrum is summed from 4 individual components
(disk, comptonization, synchrotron, and synchrotron self-comptonization
spectra), each computed by distinct routines within a single Fortran
source file.  To analyze or visualize the independent contribution
of each component it is necessary to isolate each from the total flux.
To achieve this the model can be executed in different modes, one of
which will execute {\tt write} statements to output component contributions
to disk files as they are computed.  Because our \slirp\ wrapper
for this jet model makes every routine within the Fortran file visible
to the top-level \slang\ interpreter, not just the outermost model
routine, the component fluxes can in principle now be accessed by making
the relevant routine calls, or accessing the associated common blocks,
directly from \isis.  To our knowledge the only way to achieve a similar
result in \xs\ would be to wrap the code twice, once with \initp\ to make
the custom model routine visible to the {\tt model} command, and again as
a \tcl\ extension module to expose the remaining routines to the
\tcl\ interpreter.  Tools like {\tt ftcl} or
{\tt critcl}\footnote{http://ftcl.sourceforge.net, http://wiki.tcl.tk/2523}
can simplify bindings generation in the latter case, but not automate it
to the level that \slirp\ does; it is also unclear whether they support
a range of features seen in general scientific codes, such as common blocks,
string and multidimensional arrays, or complex datatypes.

Using \initp\ for models which call external libraries can also be
problematic.  For example, collaborators are extending the above
jet model with codes that call GNU Scientific Library routines, but
\initp\ documents no mechanism by which the GSL can be linked in at build time.
Following compiler convention, this is achieved with \slirp\ by specifying
{\tt -lgsl} on the command line.

Finally, functions wrapped by \slirp\ can be automatically vectorized,
allowing them to operate over entire arrays in a single call, and at
the speed of compiled C code instead of using slower, interpreted looping
constructs.  Vectorized wrappers can also be tuned for
parallelization with OpenMP \citep{Noble.2007.OpenMP}, allowing scientists
to take better advantage of their multicore machines during \isis\ analysis.
To our knowledge autogenerated vector parallel wrappers are currently
unique to \slirp, and therefore inaccessible in \xs.

\section{Parallel and Distributed Computation}

Parallel computing is not new, but to this day few astronomers employ
parallelism for standard problems in data analysis.  To provide a 
quantitative sense of this assertion relative to X-ray spectroscopy
\citep{2006ASPC..351..481N},
we performed the following full text searches on all published papers
indexed in the ADS \citep{1993ASPC...52..132K} by September of 2005.
\begin{table}[h]
 \footnotesize
 \begin{center}
  \begin{tabular}{cc}
  Keywords & Number of Hits \\
  \hline
  parallel AND pvm & 38  \\
  message AND passing AND mpi & 21 \\
  xspec & 832 \\
  xspec AND parallel AND pvm & 0 \\
  xspec AND message AND passing AND mpi & 0 \\
  \end{tabular}
 \caption{Published references to parallelism and \xs, as indexed in ADS
    through 2005.}
 \end{center}
\end{table}
Extra keywords were included with PVM and MPI (the Message Passing
Interface, \citealt{330577}) so as to cull false matches (e.g. with
the Max Planck Institute).  Queries in ADS on other modeling tools,
or with other search engines such as Google,
all yielded similar trends: astronomers and
astrophysicists have employed parallel computing, but mainly for highly
customized, large-scale problems in simulation, image processing, or
data reduction.
Even though a majority of papers published in observational astronomy
over recent decades
result from fitting models within established software systems, virtually
no one is employing parallelism to do so, especially in the interactive
context.

As discussed in \cite{2006ASPC..351..481N} and earlier
in \S \ref{scalability}, for several years PVM has been used in
ISIS\footnote{PVM was used due its
longstanding tolerance of faults on heterogeneous clusters
of workstations, but in principle an MPI module could be used just as
easily within \isis.} to apply
pools of machines to coarse-grained calculations that do not fit
within the compute space of one workstation.
Consider relativistic Kerr disk models, for example.  Historically,
implementors have opted to use precomputed tables to gain speed at
the expense of limiting flexibility in searching parameter space.
However, by recognizing that contributions from individual radii may
be computed independently the model has been parallelized in \isis\ to
avoid this tradeoff \citep[A. Young, priv.  comm.,][]{2006ApJ...652.1028B}.
Likewise, the aforementioned jet model can be
expensive to compute, particularly when calculating error bars: generating
90\% confidence limits for 12 free parameters, at the
very coarse tolerance of 0.5, required 4 days on a single 2.66 GHz
Intel Core 2 Duo processor.
To increase performance we assigned the confidence limit search to 12
Intel Xeon 3.4GHz processors\footnote{Using new versions of the
{\tt cl\_master} and {\tt cl\_slave} scripts described on the \isis\
website.} within
a PVM.  This reduced the runtime at 0.5 tolerance to under 24 hours, a
greater than 75\% speedup; it also allowed us to discern finer features
in parameter space by increasing the tolerance resolution by a factor
of 500, to 0.001, while keeping the overall runtime to 50 hours, or
approximately half of the serial runtime at 0.5 resolution.

Temperature mapping is another problem that was straightforward to
parallelize with \isis. For instance, \cite{2004cosp...35.3997W}
provides a
map of heating in the intracluster medium of Perseus, computed from
10,000 spectral extractions and fits on 20+ CPUs in just several hours.
Additional efforts have also led to improvements in related areas of
research, such as {\tt pvm\_xstar}, a
parallelizing wrapper for XSTAR which has made it feasible for us
to probe thousands of photoionized gas physical scenarios in the time
it has previously taken to compute only a handful of such models
(Noble \& Ji, in prep.).

At the other end of the architectural scale, we have also shown
that \isis\ can make transparent use of OpenMP to exploit shared
memory multiprocessors \citep{Noble.2007.OpenMP}.  This is especially
relevant in light of the emergence of multicore architectures: soon
most astronomers will have parallel computers on their desktops, but
few astronomy applications are poised to take advantage of them.
Analysis systems which do not embrace parallelism
can process at most the workload of only 1 CPU, resulting in a
dramatic $1/n$ underutilization of resources as more CPU cores are
added.  However, astronomers are well versed in scripting, particularly
with very high-level, array-oriented numerical packages like IDL, PDL,
and \slang, to name a few.  They combine easy manipulation of mathematical
structures of arbitrary dimension with most of the performance of compiled
code, with the latter due largely to moving array traversals from the
interpreted layer into lower-level code like this C fragment
{\small
\begin{verbatim}
        case SLANG_TIMES:
             for (n = 0; n < na; n++)
                 c[n] = a[n] * b[n];
\end{verbatim}
}
\noindent
which provides vectorized multiplication in \slang.
This suggests that much of the strength and appeal of numerical scripting
stems from relatively simple loops over regular structures, making
them ripe for automatic parallelization.  This pattern was exploited
in \slirp\ and \slang\ to effect the OpenMP parallelizations
described in \citep{Noble.2007.OpenMP}.

We believe that \isis\ is the only general purpose spectroscopy system
in which such a range of parallelism---from single processors on
multiple machines to multiple processors on single machines---has
been demonstrated.  In both cases speedups were obtained without
exposing parallelism at the \isis\ application level.  For instance,
the same serial implementations of spectroscopic models have been used
for both single- and multi-processor execution.  This minimizes two
traditional barriers to the use of parallelism by non-specialists:
learning how to program for concurrency and recasting sequential
algorithms in parallel form.  We believe these considerations are
important because the emerging ubiquity of multicore architectures,
combined with the ever-growing size of datasets and analysis complexity,
makes the regular use of parallelism in astronomy not a question of
{\em if} it will occur, but rather one of merely {\em when}.

\section{Beyond 1D Spectroscopy}
\label{beyond1D}

Data from modern telescopes are getting larger and more detailed,
with the use of multiple datasets from multiple missions becoming
commonplace.  Making sense of it all often requires techniques
more sophisticated than traditional plots or images.  Canned data
reduction threads, while an important piece of the puzzle, can only
go so far.  Data openness and architectural modularity
(Fig. \ref{arch}) make it easy to push
\isis\ beyond its original mission of 1D spectroscopy to address
this broader set of analysis problems.  As we illustrate here,
it has become unnecessary to use entirely separate applications
(as XImage, XSelect, or Xronos are used alongside \xs) or be constrained
by a strictly \ftf\ model, in order to
supplement spectral modeling with, for example, advanced
filtering, imaging, or timing analysis.  For each of the examples
shown here it is not clear how a similar result might be
obtained so directly within \xs.
\subsection{Timing Analysis}
\begin{figure}[t]
  \begin{centering}
  \includegraphics[scale=0.35]{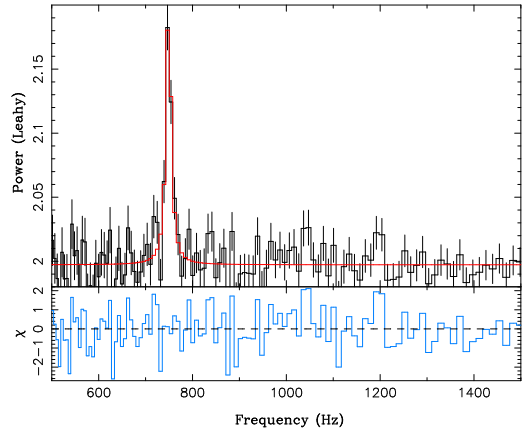}
  \caption{Power Spectrum of X1820-303, generated from the PCA in RXTE
  observation P10075 (data set 10075-01-01-02).}
  \label{timing}
  \end{centering}
\end{figure}

The optional SITAR module\footnote{http://space.mit.edu/CXC/analysis/SITAR}
endows ISIS with timing analysis capability, obviating the need
to use a separate timing application such as XRONOS. The
power spectrum in Fig. \ref{timing} was generated entirely in ISIS:
the data were read, the FFTs were performed and averaged,
the Power Spectrum was logarithmically binned over Fourier frequency, the
constant + two Lorentzians model was fit, and the results plotted, 
operating directly upon arrays in memory, all without ever leaving
the program.
\subsection{Modular Imaging with \slang\ Numerics}
\label{ds9}
Many excellent imaging tools are available to astronomers, but it is
impossible for any one of them to do everything.  Here again, the
extensibility and modularity of the \isis\ paradigm helps avoid the
fate of ``doing nothing" while waiting for the implementation of new
features desired in one's work.  In such cases it can be expedient
to craft new modules which add missing features, or make certain
functionality more easy to incorporate into modeling and analysis.

\begin{figure*}
  \begin{centering}
  \hspace*{3.5mm}
  \includegraphics[angle=-90,scale=0.35]{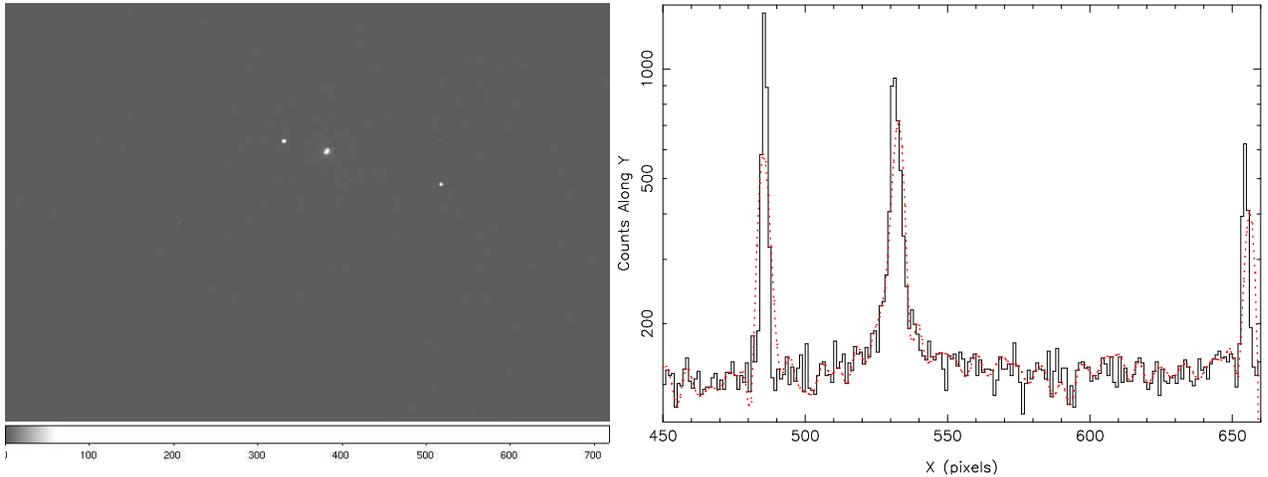}
  \includegraphics[angle=-90,scale=0.35]{f7b.eps}
  \caption{ACIS image from 2004 Chandra observation of NGC7714, with
     corresponding square and smoothed pixel intensity histograms.}
  \label{ngc7714}
  \end{centering}
\end{figure*}

For example, consider Fig. \ref{ngc7714}, drawn from Chandra observation
4800 of interacting galaxy pair NGC7714,
and the following series of commands:
{\small
\begin{verbatim}
    isis> require("ds9")
    isis> require("gsl")
    isis> image = ds9_get_array()
    isis> hist = sum(image, 0)
    isis> range = [440:680]
    isis> hplot(range-1, range, hist[range])
    isis> xa = [440:680:4]; ya = hist[xa]
    isis> smoothed = interp_cspline(range, xa, ya)
    isis> oplot(range, smoothed)
\end{verbatim}
}
First the SAOds9 and GNU Scientific Library extension modules
\citep{2005ASPC..347...17P} are loaded; we
show this explicitly for didactic purposes, but \isis\ can be configured
to load either module automatically when the given functions are invoked.
The 1024$\times$1024 image is then retrieved directly into a
properly byteswapped
2D \slang\ pixel array, using an XPA \citep{1995xpa.v1M} binary transfer 
instead of intermediate file I/O.  This image
is collapsed along its X axis using the native \slang\ { \tt sum} function,
to create an intensity histogram plot. The GSL cubic spline function
is then called to smooth the brightest portion
of the intensity histogram, interpolating over every fourth point therein,
and finally we overplot the result (dotted red line).
Although the task here is relatively straightforward, it again shows how
open-ended
analysis objectives can be achieved by weaving existing tools together 
in new ways, using an array-based interpreter as the thread.  DS9 is
extended beyond its essential role as a qualitative display tool and
into the realm of quantitative analysis, while the \isis\ user is able
to exploit the imaging capabilities of DS9 within interactive or
scripted analysis.

\begin{figure*}
  \begin{centering}
  \includegraphics[scale=0.4]{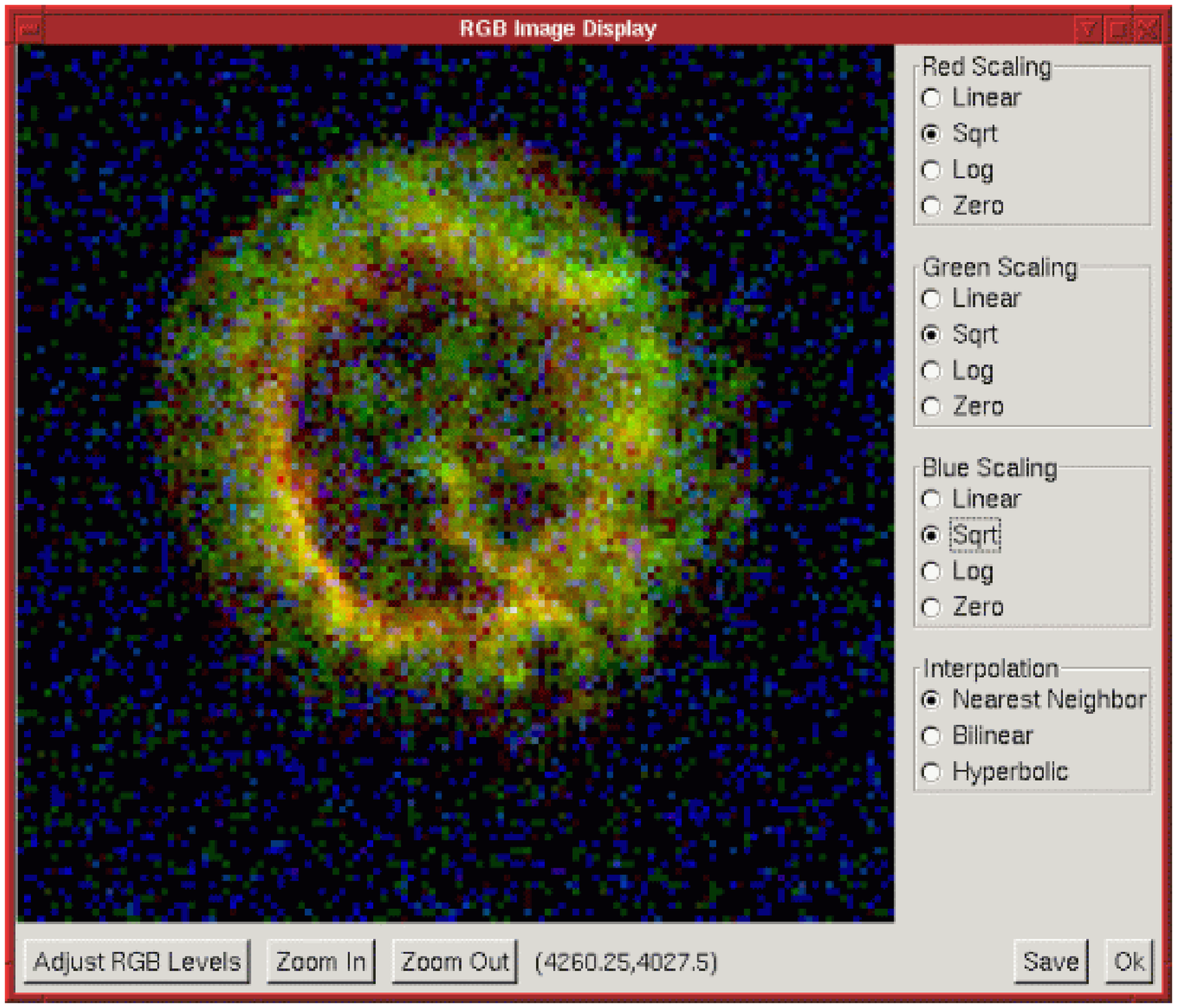}
  \hspace*{4mm}
  \includegraphics[scale=0.4]{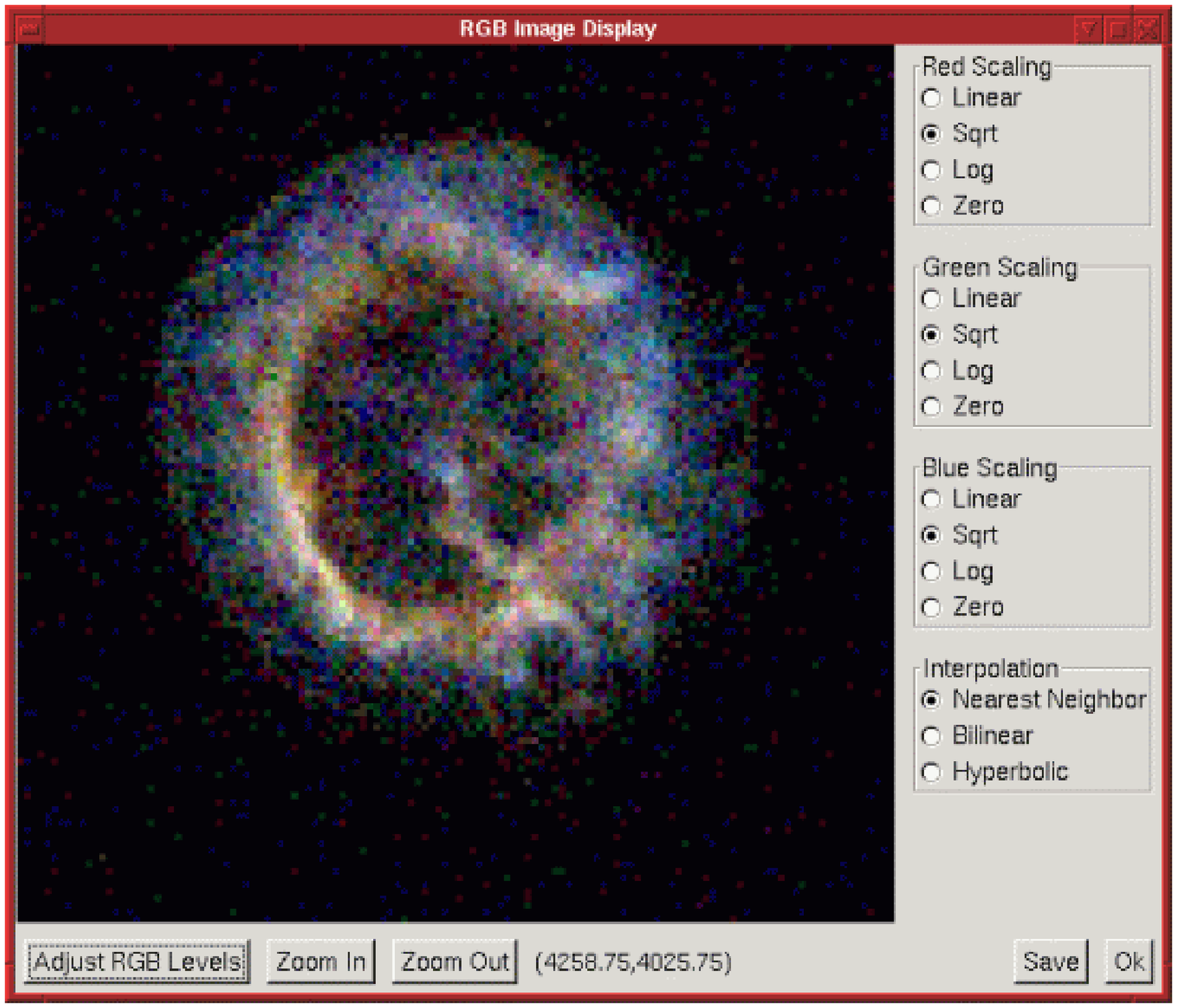}\\
  \vspace*{2mm}
  \includegraphics[scale=0.37]{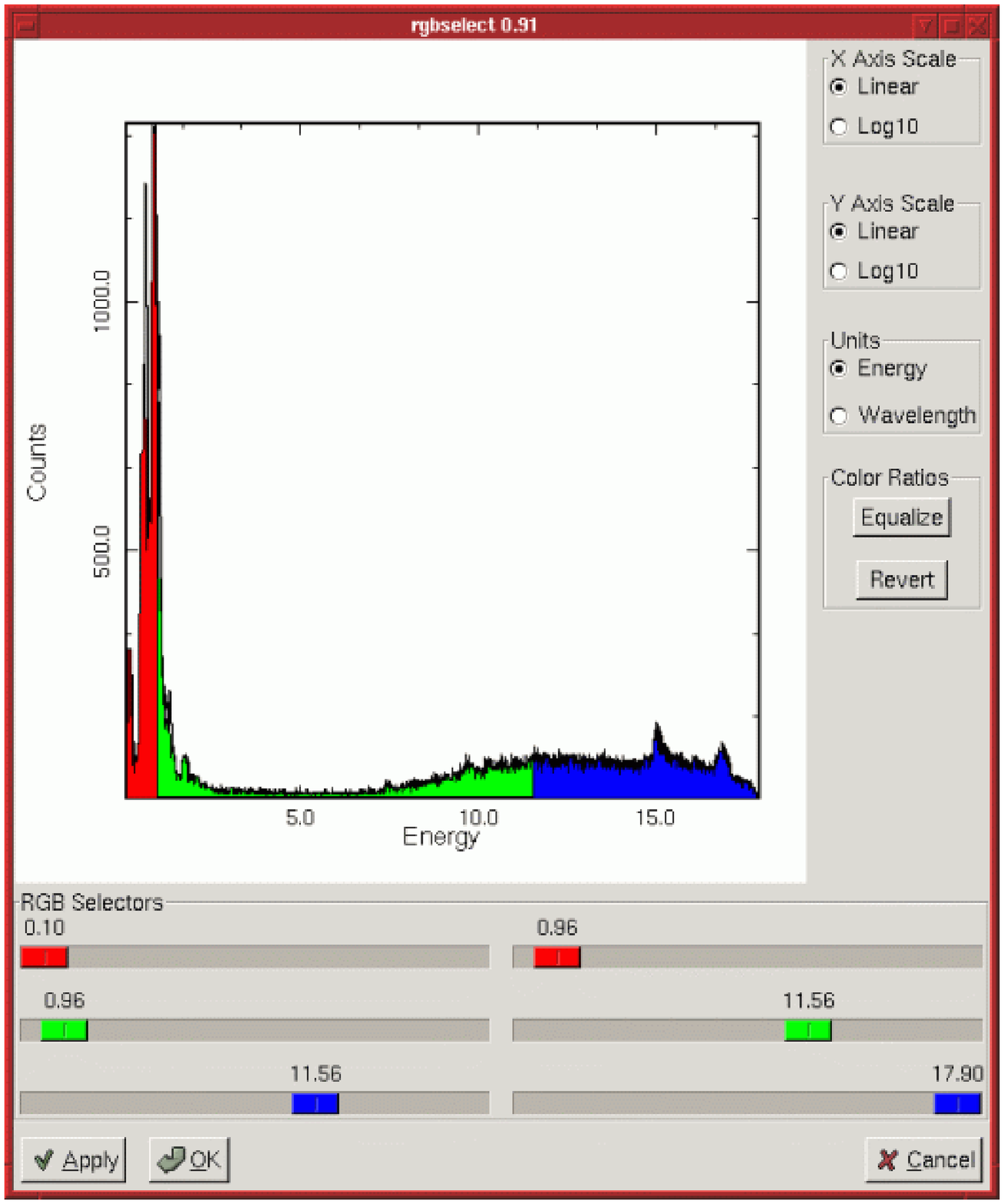}
  \hspace*{10mm}
  \includegraphics[scale=0.37]{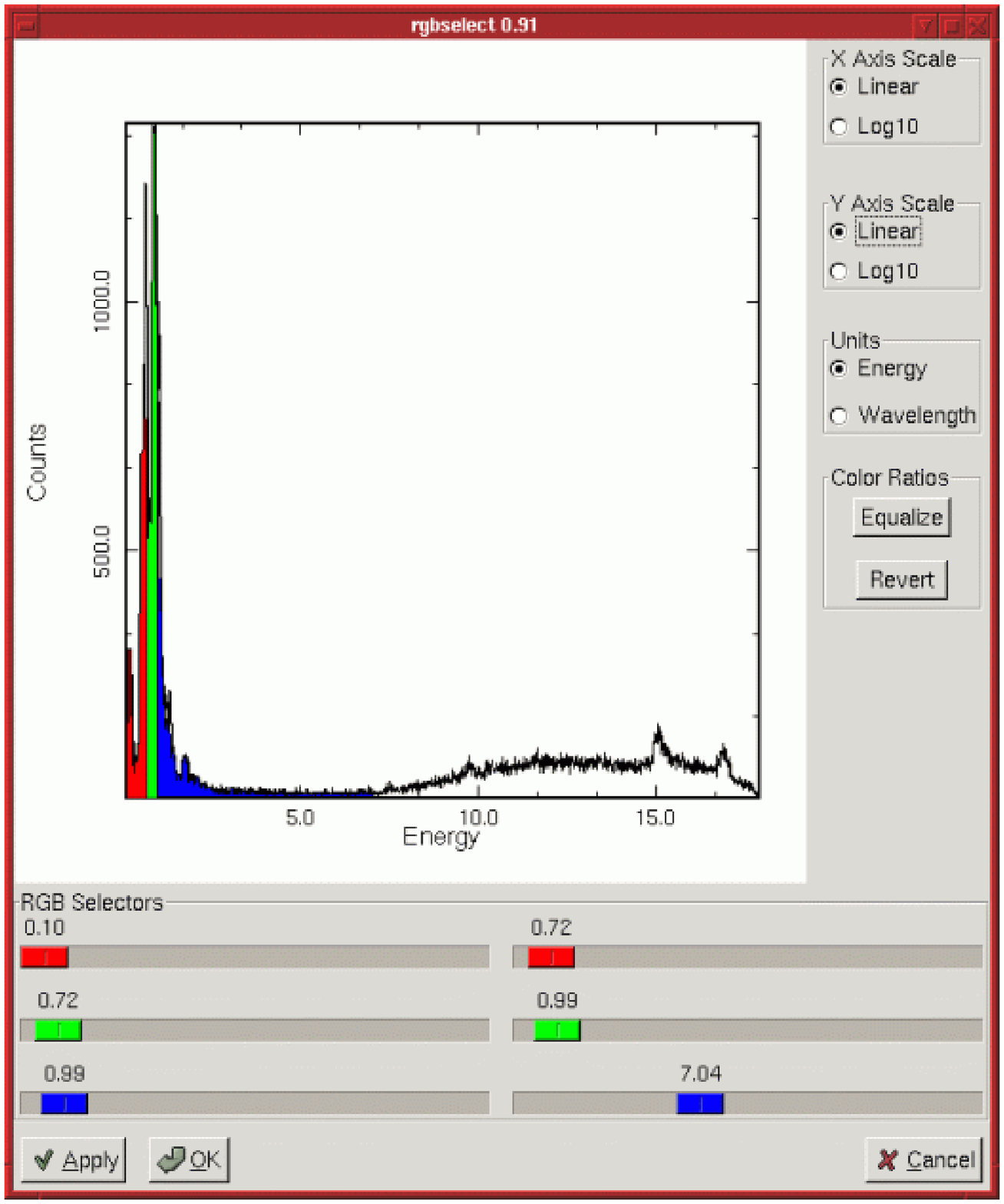}
  \caption{
  Interactive 3-color image creation in \evtimg, using the level-2 event file
  acisf00120N001\_evt2.fits from Chandra observation 120 of supernova
  remnant E0102-72.  One may pan around (by clicking to center the image),
  zoom in and out, and set various scaling properties by clicking on the
  corresponding buttons. The image in the left column shows a fair amount
  of high energy background noise, as evidenced by the blue
  background pixels.  These were removed in realtime by adjusting the
  colored energy band sliders that define the RGB levels (bottom panels),
  to yield the improved image in the right column.}
  \label{evt2img}
  \end{centering}
\end{figure*}

\begin{figure*}[h]
  \begin{centering}
  \includegraphics[scale=0.5]{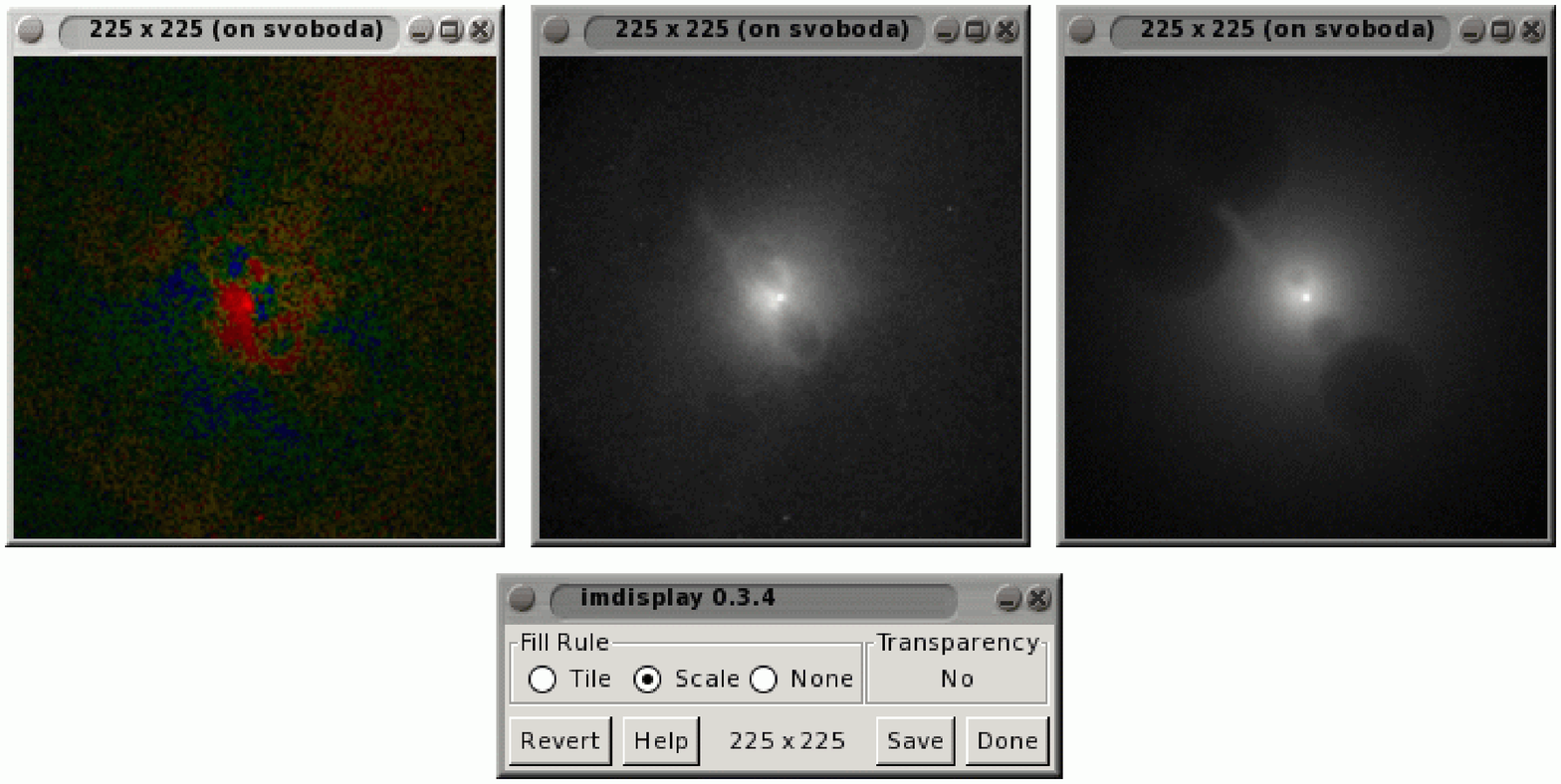}\\
  \vspace*{10mm}
  \includegraphics[scale=1.05]{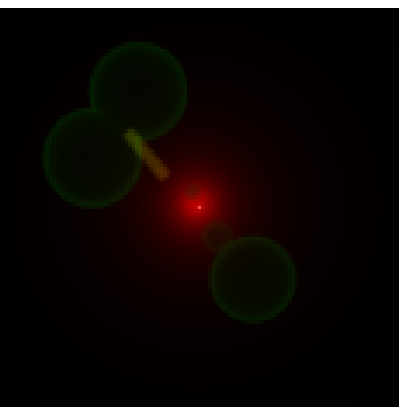}
  \hspace*{10mm}
  \includegraphics[scale=0.47]{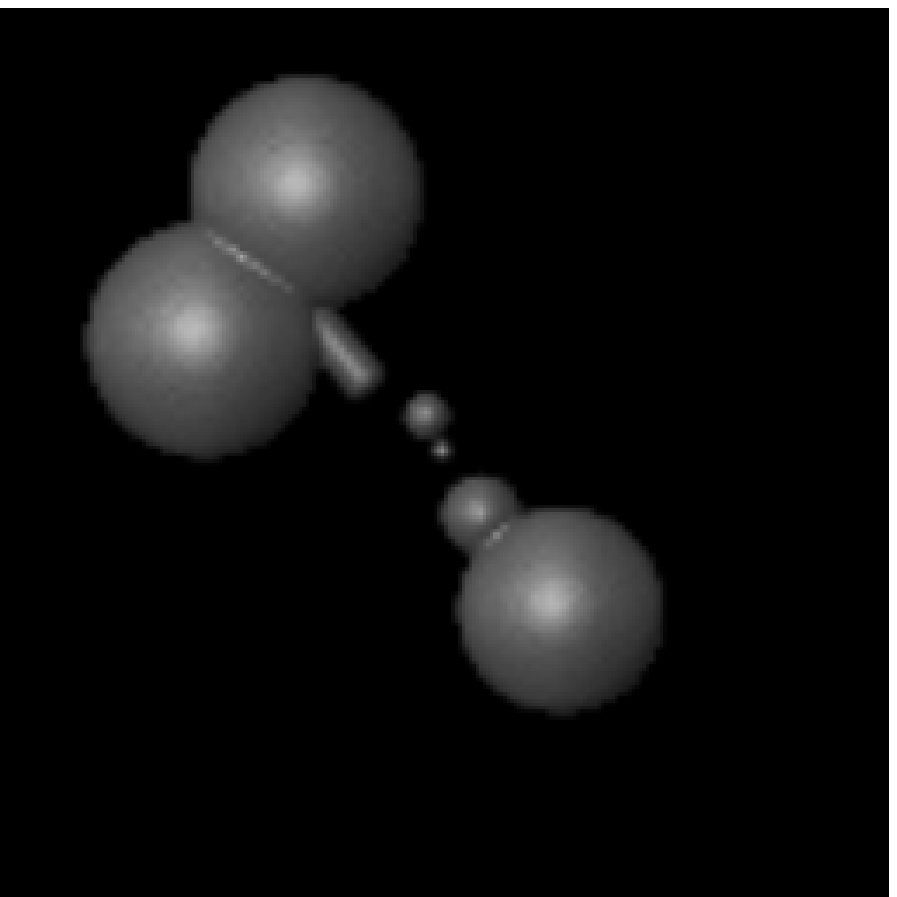}
  \caption{Images created in \isis\ by HYDRA {\tt Event-2D} and {\tt Source-3D}
  routines, while analyzing cavities within the Chandra observation 4969
  of Hydra A.  Top: residuals, data, and model.  Bottom: 2D and 3D
  projections.}
  \label{e2d}
  \end{centering}
\end{figure*}

Another example is \evtimg,\footnote{http://space.mit.edu/cxc/software/slang/modules/hist/evt2img.html} a simple guilet\footnote{This
term is used in \cite{2005ASPC..347..237N}
to refer to graphical applets embedded within a primarily command-line
application, and typically coded in a scripting language, allowing
the use of graphical interfaces when convenient while avoiding an
exclusively graphical interface.}
which combines the histogram and Gtk modules to provide an interactive
mechanism for creating 3-color images directly from event lists.
Images are created in \evtimg\ by defining 3 energy band filters, 
binning the events selected by each filter with the {\tt hist2d} function,
and overlaying the resulting red, green, and blue monochrome images 
within a Gtk display window.  As shown in Fig. \ref{evt2img}, a useful
feature of \evtimg\ is its ability to plot the energy spectrum as a 1D
histogram in order to tune the color band filters in real time, simply
by adjusting
the individual red, green, and blue sliders.  For interactive exploration
this method is faster and
more intuitive than the traditional \ftf\ approach, and produces
no temporary file litter while experimenting with color assignment filters.
{\tt Evt2img} may be run standalone from the operating system prompt or
as a function in \isis, and in the latter case allows event data to be
input directly from in-memory interpreted arrays in addition to traditional
event files.
The HYDRA project at MIT provides a wealth of more sophisticated
examples\footnote{http://space.mit.edu/hydra/E2D\_demo/e2d\_demo.html}
of how advanced imaging can be tied more directly to numerical modeling
and analysis in \isis.  Under the auspices of the NASA AISRP program, a
broad suite
of 2D and 3D fitting, geometry, and visualization routines have been
developed, e.g. to do forward folding and comparison of models with 2D
event-based data sets.  These are described in detail online,
 but are briefly illustrated
here in the context of an analysis of cavities in Hydra A, using the
Chandra observation 4969.  A model is defined by combining traditional
spectral components (e.g.  mekal and powerlaw) with 3D geometric components
(e.g. an AGN modeled as a sphere).  The fitting is performed directly in
\isis, as is the generation of the residual, data, and model images, and
2D and 3D model component projections (Fig. \ref{e2d}).
\begin{figure*}
  \begin{centering}
  \includegraphics[scale=0.70]{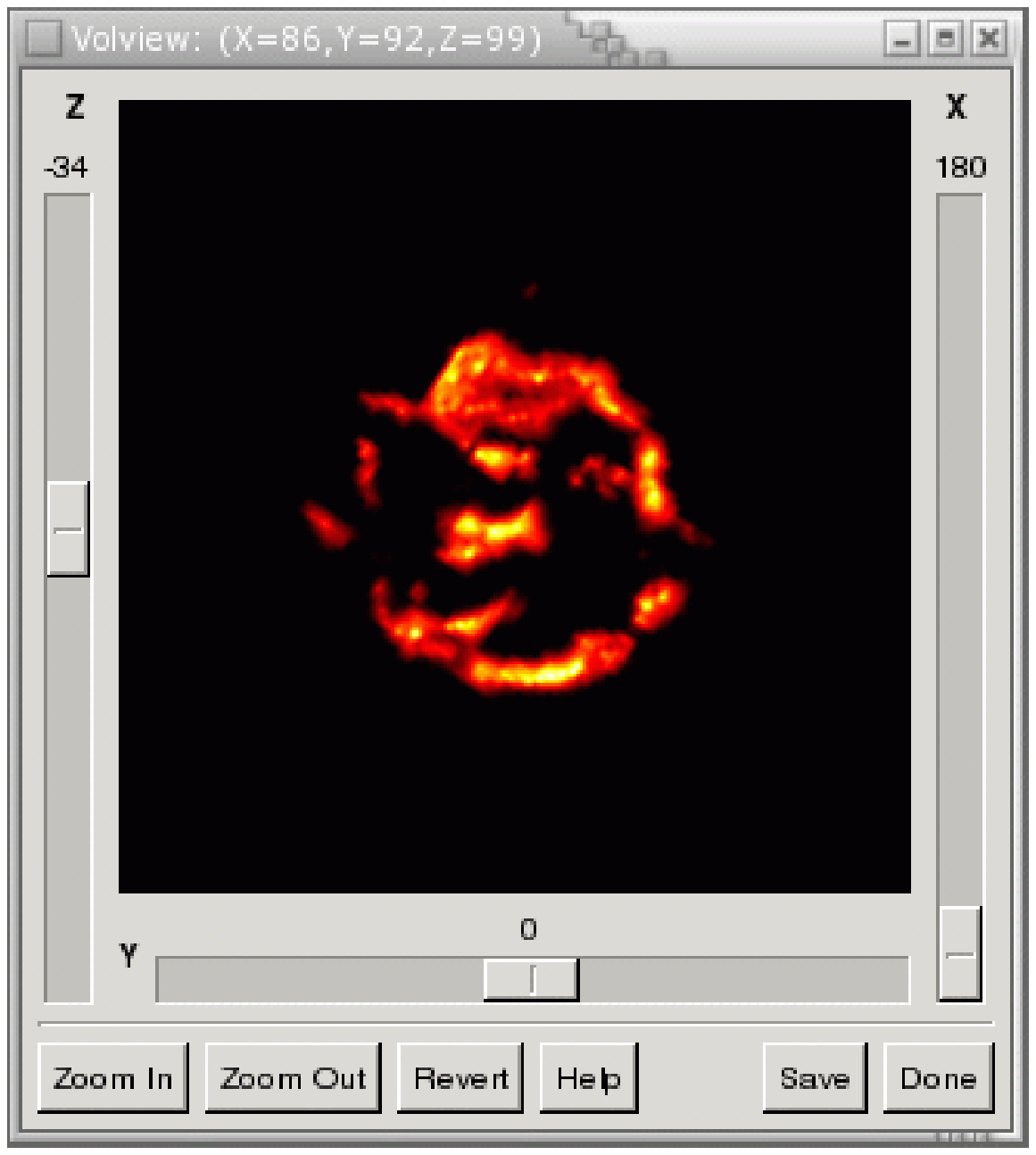}
  \hspace*{0.2in}
  \includegraphics[scale=0.70]{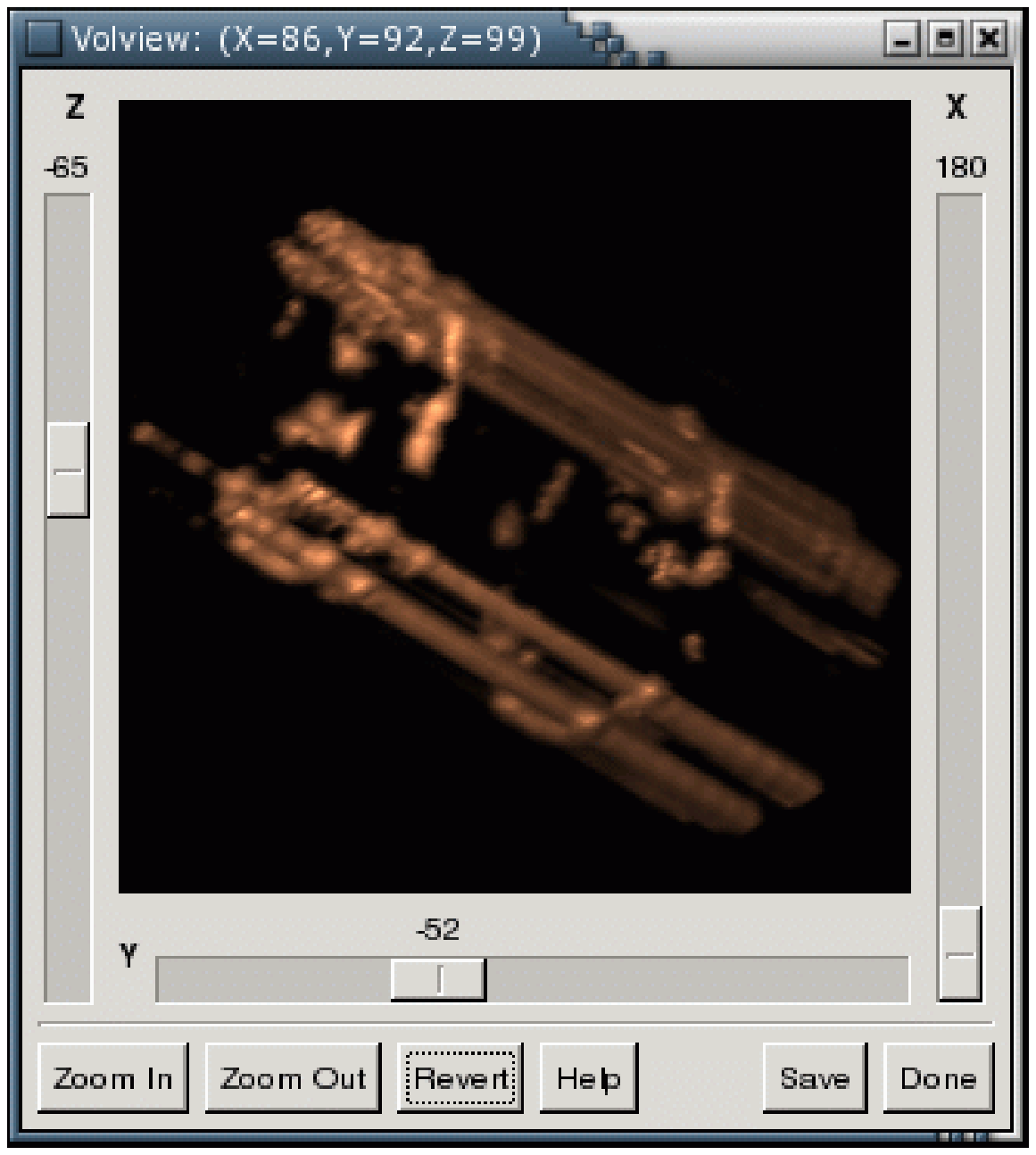}
  \caption{Entire datacube from Spitzer observation 3310; the two clumpy
	   areas are instantly recognizable as regions of strong emission.}
  \label{spitzer-volume}
  \end{centering}
\end{figure*}
\subsection{Volume Visualization}
Astrophysical observation and simulation are in the midst of a transition
beyond 1D spectra and 2D images, and into the realm of three-dimensional
phenomena. A number of astronomy software tools exist which enable
visualization of so-called 3D data cubes, but a problem shared by many of
them is that they are limited to displaying one 2D slice at a time,
optionally in series as an animation.
Here we consider a 99 spectra data cube generated by
CUBISM\footnote{http://ssc.spitzer.caltech.edu/archanaly/contributed/cubism}
from Spitzer observation 3310 of Cassiopeia A, with dimensions RA,
DEC, and $\lambda$.  To identify where emission is strongest in
the spatial and wavelength domains, the entire FITS datacube
can be passed directly to \volview\ (Fig. \ref{spitzer-volume})
{\small
\begin{verbatim}
  isis> require("volview")
  isis> cube = fits_read_image("casa_ll1_s12.fits")
  isis> volview(cube)
\end{verbatim}
}
\noindent
This visualization makes it instantly clear that the clumpy regions
correspond to bright emission at fairly specific wavelengths.
The \volview\ guilet
is our
interface to the Volpack
rendering library \citep{192283}, which, although somewhat dated,
is small, has no external dependencies, requires no special
hardware, and is very fast.  The shear-warp factorization algorithm
in Volpack has been generally recognized as the fastest rendering
technique available \citep{353903}, a crucial factor for interactive
analysis.  In contrast to high-end tools like ParaView and
VISIT\footnote{http://www.paraview.org,\hspace*{2mm}http://www.llnl.gov/visit},
volview provides a simple, low buy-in path to volume rendering and
is actively used in the HYDRA project.\footnote{http://space.mit.edu/hydra/v3d.html}.

\subsection{Visual Correlation and Multidimensional Filtering}
Having the ability to cut, visualize, and correlate data, in complex
ways and from multiple missions, is invaluable to analysis.  This
exploratory process ultimately seeks to derive sets of constraints
{\em \bfseries C} = \{$C_{0}, ..., C_{n}$\} upon input data
{\em \bfseries D} by iterating through a series of models, comparisons
to data, and refinements of assumptions and model parameters.
Because its analytic process does not encompass whole-array
numerics (\S \ref{numerics}, \S \ref{input}), an iteration of this cycle
in \xs\ can involve dumping application state to disk, followed
by the execution of multiple tools such as {\tt fselect} or \fcalc\
to transform or cut file data, possibly \fv\ or XIMAGE to visualize
% NB: fv cannot filter table columns by laying regions down upon plots,
% nor can data "shrunk" by the table calculator [e.g. where(x > a) is
% not possible, b/c where() doesn't exist and b/c the fv calculator is
% table-based, and so operates upon whole columns with #rows kept constant
filtered data subsets, and \xs\ to incorporate the results back into analysis.
This process creates temporary file litter products which can quickly
become difficult to manage, and requires slow disk I/O passes over files
when applying each unique set of filter constraints.  Other limiting
factors are that it restricts the scope of exploration to data expressed
in the FITS file format, and that \tcl-based I/O in tools like \fv\ does
not scale well to datasets containing millions of events.

\label{vwhere}
In contrast, \vwhere\ \citep{2005ASPC..347..237N}, a graphical extension
of the \slang\ { \tt where} array slicing function, unifies the constraint
cycle by enabling each phase to be performed within a single
application.  The result can be an intuitive, faster, and
more powerful alternative to file-based data mining. 
By supporting the use of interpreted \slang\ arrays \isis\ \& \vwhere\ make
it easy to combine data from multiple sources and formats, not just
FITS files, and examine them in natural ways, as well as foster
scalability.  \vwhere\ has been used to
simultaneously mine hundreds of observations from multiple telescopes,
e.g. for evidence of transitions between the soft and hard states of
Cygnus X-1 \citep{2005ApJ...626.1006N}, correlate model parameters with
fit results---directly in memory from dozens of fits---for scores of
observations \citep{2006A&A...447..245W, MarkoffM81InPrep, NowakCygX1InPrep},
and quickly corroborate colleague results \citep{2007ApJ...663L..93H}.

\begin{figure*}[p]
  \begin{centering}
  \includegraphics[scale=0.6]{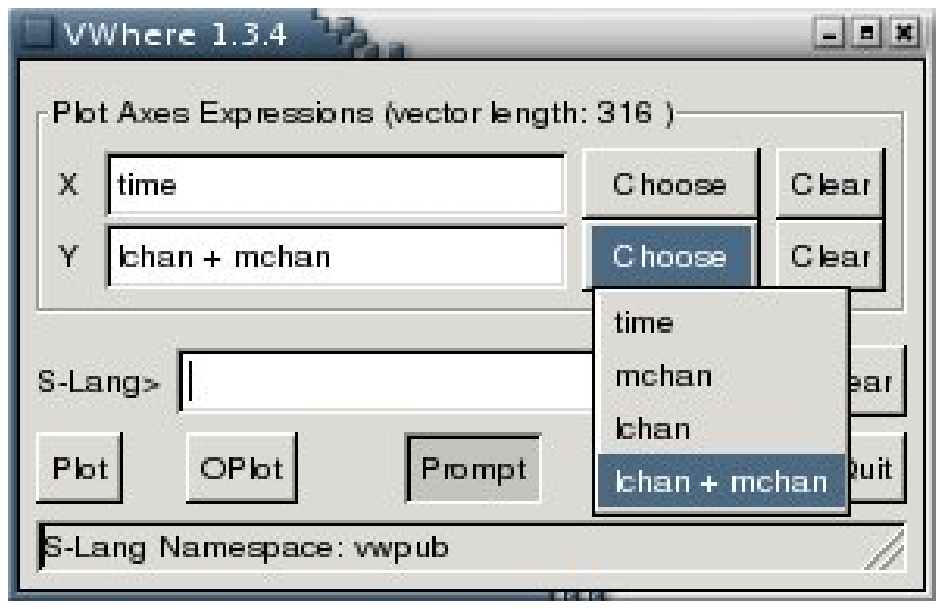}
  \hspace*{0.51in}
  \includegraphics[scale=0.4]{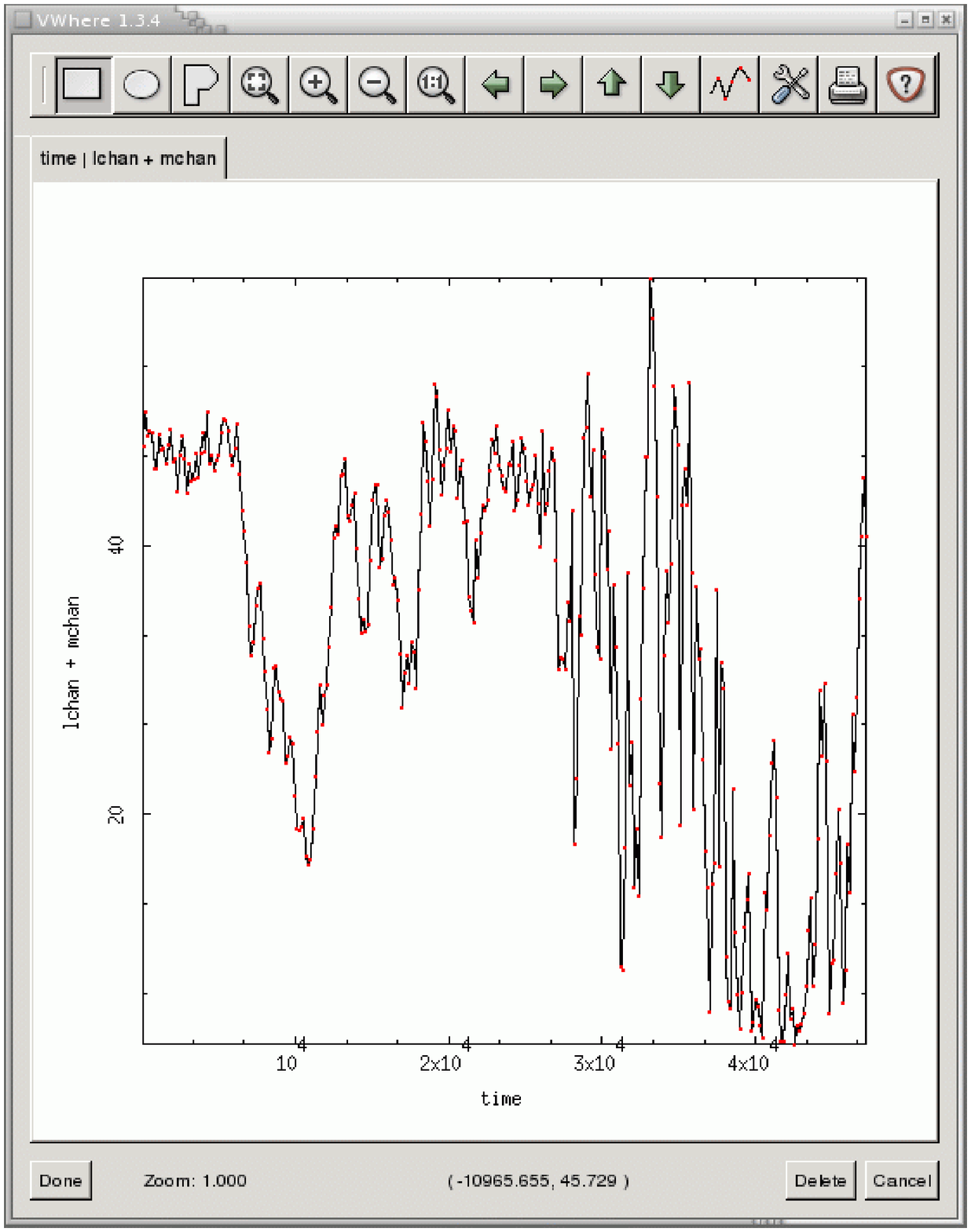}
  \caption{\vwhere\ in \isis\, using SITAR and the AGLC package to
     generate a lightcurve for Chandra observation 3814 of Cygnus X-1.
     Configurable plots are generated from an axis expression window
     containing several text fields.  Each X/Y expression may contain
     any valid \slang\ statement, including calls to C, C++ or \fort\ 
     functions within external modules.  The chief constraint upon an
     expression is that, when evaluated, it generate a numeric vector
     equal in length to the existing data vectors.
     Here we show \vwhere\ launched with a struct
     containing 3 input vectors, as well as how easily new vectors (e.g.
	{\tt lchan + mchan}) can be fabricated from them on the fly.
     Each vector expression is saved in the {\tt Choose} dropdown menu,
	for easy reuse with no retyping.
     Vectors of arbitrary size may also be overplotted, while the 
     {\tt S-Lang>} prompt may be used as an interactive command line,
	e.g., to issue \isis\ commands or execute arbitrary \slang\ code.}
  \label{vwhere1}
  \end{centering}
\end{figure*}

\begin{figure*}
  \begin{centering}
  \includegraphics[scale=0.4]{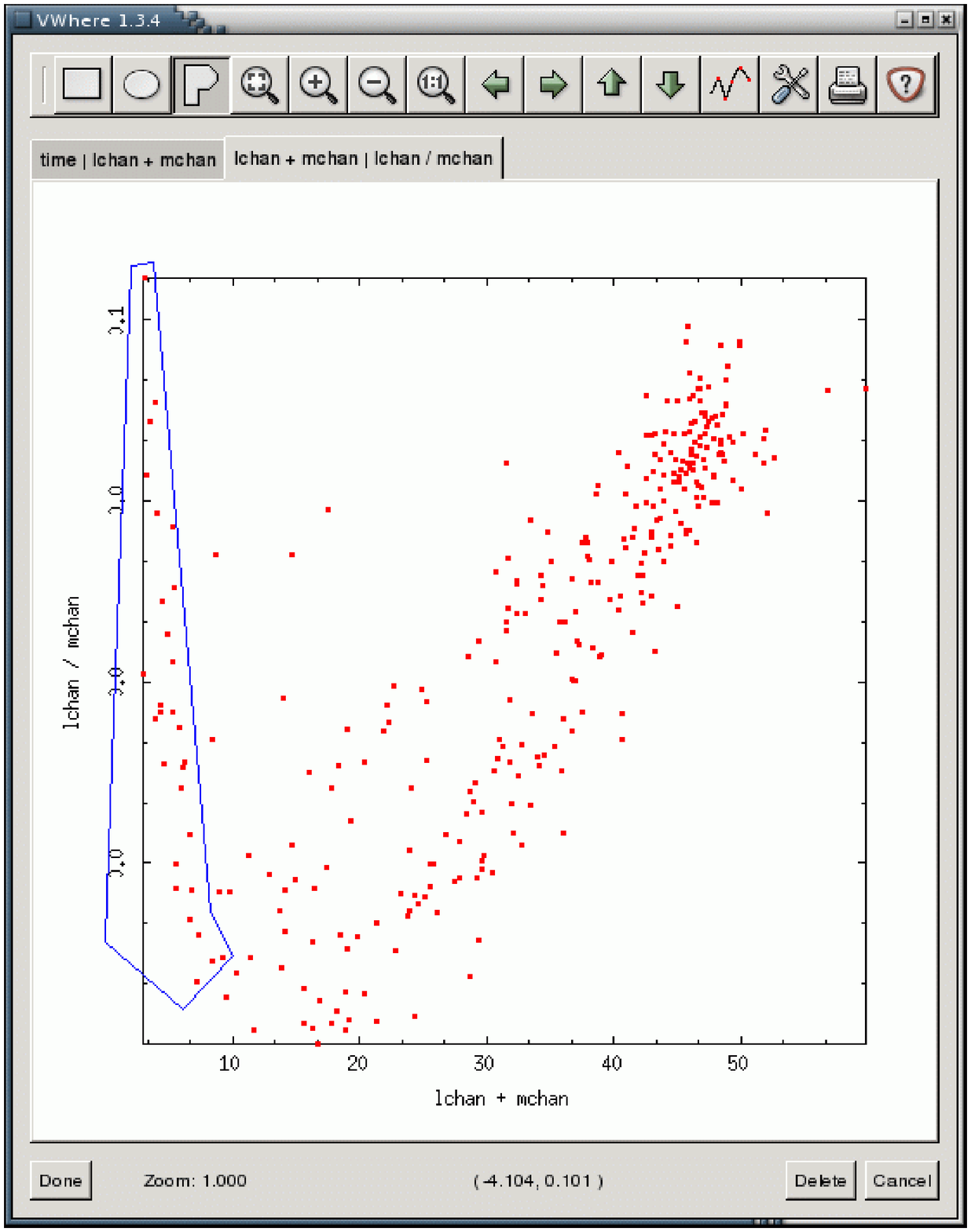}
  \hspace*{0.1in}
  \includegraphics[scale=0.4]{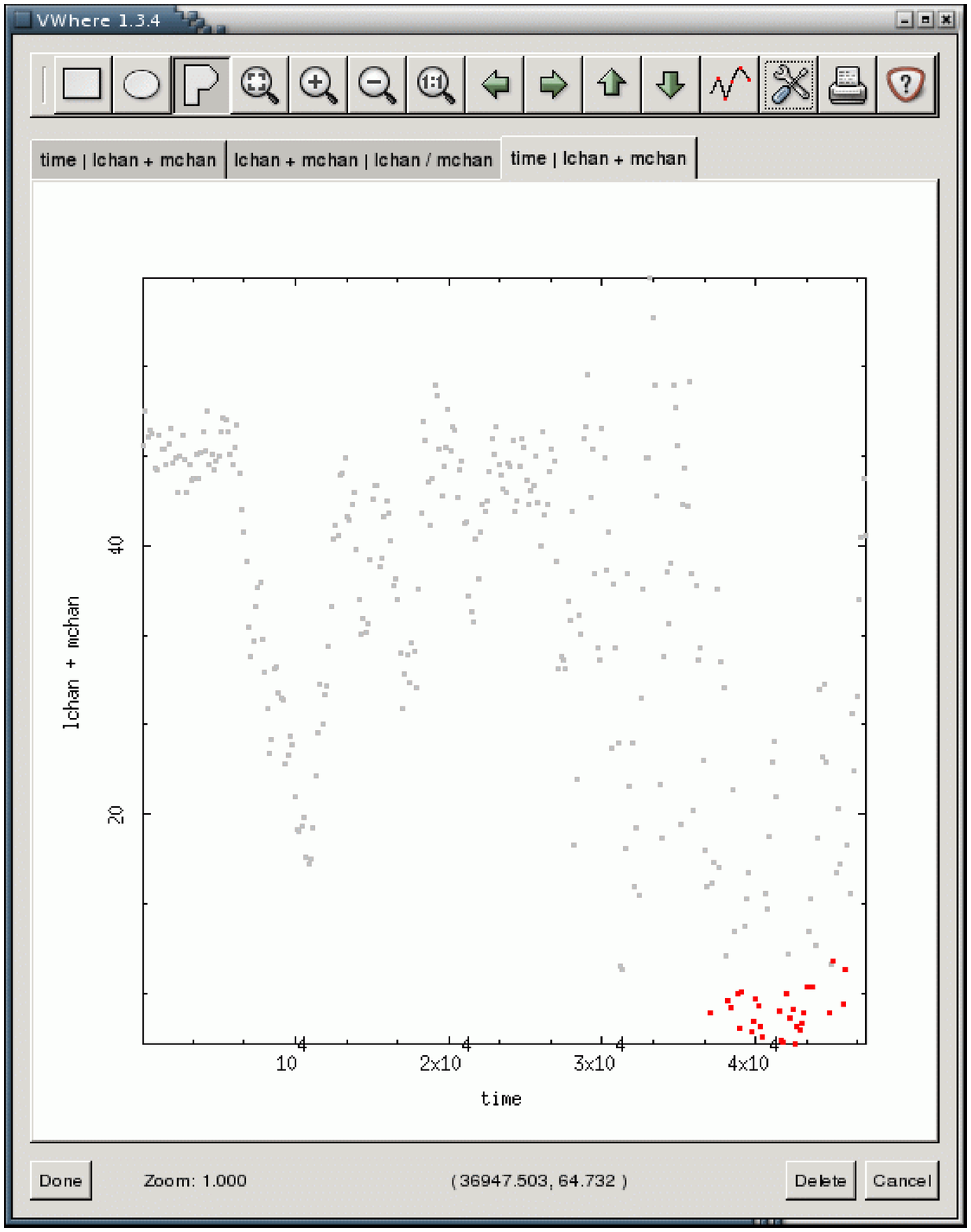}
  \caption{A polygon filter applied to the color intensity diagram, and
  its effect upon the lightcurve  (Hanke et al, in prep.).}
  \label{vwhere2}
  \end{centering}
\end{figure*}

Filtering in \vwhere\ amounts to manipulating regions of interest
on plots generated from \slang\ arrays (Fig. \ref{vwhere1}), with
no syntax required and the effects upon any combination of vectors
automatically visualized for inspection (Fig. \ref{vwhere2}).
This approach reveals data correlations that can be difficult and
time-consuming to ferret out with tool-based techniques.  New data
vectors may be created on the fly and with no additional I/O overhead,
using anything from simple arithmetic combinations of existing vectors
to any mathematical transformation that can be specified in \slang\
or imported from external C, C++, or \fort\ codes.

In contrast, file-based filtering tools require the use of syntax which
may conflict in subtle ways from package to package.  Specifying
mathematical operations to such tools, e.g., in the form of
quoted Unix shell syntax, can be tedious and cumbersome, and does not
approach the expressive power (consider recursion, for example) or
performance of \isis\ numerics.
As an illustration, consider the use of calculator tools for
the discriminant computation in \S \ref{numerics}: the 1-million
element \slang\ array datapoint in Fig. \ref{idl-plot} corresponds
to a runtime of ca. 0.15 seconds:
{\small
\begin{verbatim}
   isis>  a = [1:1e6+1];  b = 3*a; c = 2*a
   isis> tic; d = sqrt(b^2 - 4*a*c); toc
   0.151172
\end{verbatim}
}
\noindent
Using tools to perform similar operations upon files
{\small
\begin{verbatim}
   fcalc in out d "sqrt(b^2 - 4*a*c)"
   ftcalc in out d "sqrt(b^2 - 4*a*c)"
   dmtcalc in out expression="d=sqrt(b^2 - 4*a*c)"
\end{verbatim}
}
\noindent (from a locally compiled -O2 distribution of HEASOFT 6.1 and
a CIAO3.3 binary distribution, respectively) was consistently 1 to 2
orders of magnitude slower: roughly 1.8 sec for \fcalc, 2.3 sec for
{\tt ftcalc} and 70 sec for {\tt dmtcalc}.  If we include the time to write
{\small
\begin{verbatim}
   isis>  s = struct{a,b,c}; s.a=a; s.b=b; s.c=c
   isis>  fits_write_binary_table("in","arrays",s)
\end{verbatim}
}
\noindent the arrays to disk and then read them back in
{\small
\begin{verbatim}
   isis>  d = fits_read_table("out").d
\end{verbatim}
}
\noindent the performance penalty of the tool approach is even greater.
Moreover, the output product of each tool is not directly useful:  to
be visually inspected or used in
further analysis, for instance, it first needs to be loaded into another
program (e.g. \fv, or \xs), incurring additional I/O and
application overheads that are not seen in \isis\ because its
result arrays may be immediately manipulated or passed to subsequent
computations.

% time ftcalc in shmem://h1 d "sqrt(b^2 - 4*a*c)"
% real    2m45.378s	user    1m40.602s	sys     1m4.659s
% Fri Aug 31 14:49:22 EDT 2007
% time ftcalc in shmem://h1 d "sqrt(b^2 - 4*a*c)"
% real    2m45.787s	user    1m40.132s	sys     1m5.536s
Shared memory can, in principle, mitigate the disk I/O cost,
but in this case it did not help:  instructing {\tt ftcalc}
to write to {\tt shmem://h1} consistently resulted in runtimes of
nearly 3 minutes, while the CIAO dmtcalc tool would not permit the
creation of a shared memory FITS table.  Even if shared memory
were faster than files in this case, the point raised earlier in
\S \ref{input} still remains, namely that \xs\ documents no
clear provision though which such generic data arrays may be utilized
in analysis.
It must also be noted that few, if any, file-based transformation
tools are extensible.  If we wished to use a hypergeometric function in
\vwhere\ it is as simple as loading the GSL module with \verb+require("gsl")+
and calling the relevant function.
If the mathematical operations required for their research are not
supported, tool users would have to either make an enhancement request
and wait or create their own solution.

\bfi{
\subsection{Aside On Reproducibility}

We have been refining the approach to analysis espoused in this paper for
much of the past decade.  During that time, one of the more persistent
concerns we have privately encountered is
that it can lead to diminished reproducibility, particularly in contrast
to the \ftf\
model.\footnote{This is relevant to the overall paper because it is
the dominant means through which data are prepared for analysis in \xs,
but applies to other analysis packages as well.}
Reproducibility is a cornerstone of science, and remains a
topic of debate in wider circles \citep{HCollins.1992, 2006Natur.442..344G},
but is not treated in depth within the astronomy literature.  We
do not attempt to fill that gap here.  Rather, we aim only to address the
perception that configurable, interpreter-based methods compromise
reproducibility relative to file-based tool methods, and hope that
this may contribute to a more
complete treatment of reproducibility elsewhere.

Central to reproducibility is the recording of history.  Many tools
assist that process by annotating modified files with FITS history
keywords.  This is of clear utility, especially when tracing pipelines
and other well-defined data reduction procedures.  For open-ended
analysis, however, we do not believe it is superior to the forms of
history recorded by full-fledged analysis applications.
Consider what is being captured in history records written by tools: each
keystroke of the command used to invoke the tool.  This capability is
not unique to tools.  \xs\ and \isis\ record keystroke history by virtue
of their GNU readline command recall, editing, and logging capabilities,
as do many other systems.  Focusing on
keystrokes alone, though, obscures a larger point:  regardless of whether
one prefers logs or header keywords, indiscriminately long lists of
commands typed are of little value without some sense of their
relevance to publishable results.

In the end what matters most is that results may be regenerated so that
conclusions may be plausibly confirmed by others, rather than having
every bump or wrong turn along the way reproduced in high fidelity.
Scripts, the
logical endpoint of our interpreter-based approach to analysis, confer
this conceptual prioritization by making explicit the data and algorithms
of greatest significance.  Such scripts arise in tool-based analysis as 
well, only they tend to be expressed in system-oriented languages like
Bourne shell or Perl, rather than intrinsically whole-array numeric
languages like \slang\ or IDL.  It can be argued, then, that scripts lead
to higher forms of duplicability than FITS history records alone.  They
are also easier to annotate with additional commentary.  We therefore
conclude that that tool- and interpreter-based approaches to analysis
are approximately equivalent in the degree to which they facilitate
reproducibility.
Care must be still taken when ``feature creep" introduces incompatibilities
that make the use of older scripts problematic with newer software
versions, so it is important that version information be recorded and
that older software remains accessible on the internet.
}

\section{Conclusion}

Progress in science can be measured by our ability to pose increasingly
advanced, open-ended questions and address them with flexible techniques
of commensurate sophistication.  Since the age of Newton, one implication
of this is that scientists must also practice mathematics, and since the
age of Turing it has also meant they must practice programming.

In this paper we have compared two open-source spectral analysis applications,
\xs\ and \isis, in an attempt to gauge how their scientific and mathematical
reach may be extended with custom programming.  Contrasts between the two
have been drawn from the context of current research efforts utilizing high
performance computation, numerical modeling, atomic physics, visualization,
automated code generation, and data management.  We have
demonstrated how these have led to the incorporation of new analytic
techniques in \isis, in ways that are either unexplored or problematic
for the current \xs\ architecture and its associated \ftf\ model of
analysis.

\xs\ has been of tremendous value to the X-ray community, continues
to be actively developed, and has a strong history of community enhancement
via local models (and to some extent, \tcl/TK scripts).  However, we have
argued that the top-level simplicity of
the \xs\ interface, long and rightfully one of the pillars of its appeal,
can shroud much of its internal computational and data handling mechanisms.
This in turn can render \xs\ less adaptable---by the typical user---for
evolving research needs.  Commands which operate as black-boxed common
denominators of analysis, and permitting only file-based data input, may
not be enough to probe some of the more computationally challenging
problems facing astronomers.

We conclude that analysis applications such as \isis, endowed with
interpreted whole-array numerical capabilities and an open, modular,
and scriptable architecture designed expressly for high configurability,
are more favorably equipped to support ad-hoc research needs while not
appreciably compromising reproducibility.  Originally envisioned as a
tool for Chandra gratings spectroscopy, \isis\ has heavily influenced
the development of additional Chandra analysis software, while also
receiving NASA AISRP funding to continue its evolution within the
aforementioned HYDRA project.
Although some of the \isis\ capabilities we've emphasized do exist in
other astronomy software, we are unaware of any publications demonstrating
how similar breadths of flexibility and computational power have been
collected under the umbrella of a single open-source analysis application
and brought to bear on published research in spectral analysis and
X-ray astronomy in the manner discussed here.

\acknowledgments
We would like to thank Daniel Dewey, J\"{o}rn Wilms, and our MIT
colleagues for reviewing drafts of the paper, as well as Tracey DeLaney
for supplying the Spitzer 3D spectral cubes of CasA.  We are grateful
to our anonymous referee for constructive criticism, and to NASA for
funding this work through the AISRP grant NNG06GE58G.

\bibliographystyle{apalike}

\clearpage

\clearpage

\clearpage

%% If the table is more than one page long, the width of the table can vary
%% from page to page when the default \tablewidth is used, as below.  The
%% individual table widths for each page will be written to the log file; a
%% maximum tablewidth for the table can be computed from these values.
%% The \tablewidth argument can then be reset and the file reprocessed, so
%% that the table is of uniform width throughout. Try getting the widths
%% from the log file and changing the \tablewidth parameter to see how
%% adjusting this value affects table formatting.

%% The \dataset{} macro has also been applied to a few of the objects to
%% show how many observations can be tagged in a table.

\end{document}